\documentclass[
reprint,
superscriptaddress,
nofootinbib,
 amsmath,amssymb,
pra,
]{revtex4-2}

\usepackage{graphicx}% Include figure files
\usepackage{dcolumn}% Align table columns on decimal point
\usepackage{bm}% bold math
\usepackage{xcolor}

\usepackage{stmaryrd}
\usepackage{times}

\usepackage{todonotes}

\begin{document}

\preprint{APS/123-QED}

\title{Non-Monotonic Dynamical Correlations Across The Glass Crossover}

\author{Corentin C.~L.~Laudicina}
\affiliation{Soft Matter \& Biological Physics, Department of Applied Physics,
Eindhoven University of Technology, P.O. Box 513, 5600MB Eindhoven, The Netherlands
}
\author{Ilian Pihlajamaa}
\affiliation{Soft Matter \& Biological Physics, Department of Applied Physics,
Eindhoven University of Technology, P.O. Box 513, 5600MB Eindhoven, The Netherlands
}

\author{Liesbeth M.~C.~Janssen}
\affiliation{Soft Matter \& Biological Physics, Department of Applied Physics,
Eindhoven University of Technology, P.O. Box 513, 5600MB Eindhoven, The Netherlands
}
\affiliation{Institute for Complex Molecular Systems, Eindhoven University of Technology, P.O. Box 513, 5600MB Eindhoven, The Netherlands}

\author{Thomas Voigtmann}
\affiliation{Institut f\"ur Frontier Materials auf der Erde und im Weltraum, Deutsches Zentrum f\"ur Luftund Raumfahrt (DLR), 51147 K\"oln, Germany}
\affiliation{Department of Physics, Heinrich-Heine-Universit\"at D\"usseldorf, Universit\"atsstraße 1, 40225 D\"usseldorf, Germany}

\author{Tommaso Rizzo}
\affiliation{Institute of Complex Systems (ISC) - CNR,
Rome unit, P.le A. Moro 5, 00185 Rome, Italy}
\affiliation{Dipartimento di Fisica, Sapienza
Universit\`a di Roma, P.le Aldo Moro 5, 00185 Rome,
Italy}
\affiliation{Istituto
Nazionale di Fisica Nucleare, Sezione di Roma I,
P.le A. Moro 5, 00185 Rome, Italy}

\date{\today}

\begin{abstract}
The dramatic slowing down of structural relaxation in supercooled liquids is accompanied by the emergence of dynamic heterogeneity. A monotonically increasing dynamical correlation length, measured at the $\alpha$-timescale, is one of the remarkable features of this phenomenon. Here we show that this picture is incomplete: the dynamical correlation length measured in the $\beta$-relaxation regime exhibits a striking non-monotonic temperature dependence, reaching a maximum near the mode-coupling crossover temperature $T_c$ and decreasing upon further cooling, even as local dynamical fluctuations continue to intensify. This behavior suggests a crossover from spatially extended, maximally cooperative motion near $T_c$ to increasingly compact and localized relaxation events below it. We demonstrate that this evolution is quantitatively captured by stochastic beta-relaxation theory, an extension of mode-coupling theory beyond mean-field that explicitly predicts an avoided dynamical transition in finite dimensions. Our results provide the first direct spatial evidence in favor of the avoided-transition picture of the mode-coupling crossover, and establish the peak of the $\beta$-regime correlation length as a robust indicator of the mode-coupling crossover.
\end{abstract}

\maketitle

\section{Introduction}

To this day, a unified microscopic description of the emergent spatiotemporal heterogeneity of structural relaxation close to the glass transition remains elusive \cite{berthier2011dynamical, royall2015role, tanaka2019revealing}. As a liquid is steadily driven into the supercooled regime, its dynamics not only become extraordinarily slow, but particle motion also organizes into increasingly correlated regions of high and low mobility. This phenomenon, known as dynamic heterogeneity \cite{berthier2011dynamical}, is remarkably robust across a wide range of disordered systems. Closely analogous spatiotemporal heterogeneity has been observed not only in molecular glass formers, but also in dense colloidal and granular assemblies \cite{dauchot2005dynamical, candelier2009building} and even in biological tissues \cite{angelini2011glass}. In the context of supercooled liquids, dynamic heterogeneity
has long been linked to several universal hallmarks of glassy dynamics such as the breakdown of the Stokes–Einstein relation \cite{rossler1990indications, ediger2000spatially, swallen2003self, xu2009appearance} and the emergence of excess wings in the relaxation spectra \cite{guiselin2022microscopic}. A long-lasting open question is whether the observed phenomenology upon cooling reflects the influence of an underlying phase transition, or whether it arises from an entirely different mechanism \cite{,garrahan2002geometrical, royall2015role, hocky2012growing, dyre2026physics}

Indeed, dynamic heterogeneities usually become more pronounced around the so-called crossover temperature $T_c$ borrowed from the mode-coupling theory of the glass transition (MCT) \cite{gotze2009complex}. This reference temperature is commonly understood to mark a qualitative change in relaxation dynamics \cite{torre2004structural, flenner2013dynamic}. Above $T_c$, the structural relaxation time grows in a super-Arrhenius fashion upon cooling, consistent with the power law predicted by MCT \cite{gotze2009complex} and, more broadly, by mean-field theories of the dynamical glass transition such as replica theory \cite{parisi2020theory} and the dynamical mean-field theory \cite{maimbourg2016solution}. These theories all predict algebraic divergences of both the $\alpha$-relaxation time and of a dynamic correlation length at $T_c$, signalling the transition to a dynamically arrested state. This transition is, however, not observed in physical dimensions, where it is typically said to be \emph{avoided} \cite{kirkpatrick1989scaling}. Instead, as dynamic heterogeneities become more prominent, relaxation times typically cross over to a weaker Arrhenius-like growth\textemdash{}the precise functional form of which remains debated \cite{brambilla2009probing, mallamace2010transport, schmidtke2012boiling, tyburski2025observation, simon2026molecular}. The underlying physical mechanism of this ergodicity-restoring crossover and the extent to which it retains the character of the mean-field critical scenario, if any, remains one of the central questions of glass physics. If the crossover indeed reflects an avoided mean-field transition, it should leave quantifiable spatiotemporal imprints on the dynamics near $T_c$. This has motivated an extensive body of work on correlation lengths in glass-forming liquids.

Over the past decades, a rich variety of correlation lengthscales has been proposed in the glass literature, ranging from static ones such as the point-to-set correlation length \cite{montanari2006rigorous, biroli2008thermodynamic} to dynamic ones \cite{dasgupta1991there, lacevic2002growing, donati2002theory, tahaei2023scaling} and, more recently, lengthscales inferred from machine-learning approaches \cite{bapst2020unveiling, jung2023predicting, jung2024dynamic, jung2025roadmap}. 
To probe dynamic heterogeneity, one typically considers a family of dynamic correlation lengths, denoted $\xi_{\mathrm{d}}(t)$, which measure the range over which a local perturbation affects the surrounding dynamics. Several definitions of $\xi_{\mathrm{d}}(t)$ exist, but a key  common feature is its explicit time dependence: it is expected to grow during the correlated stage of relaxation and recede once structural relaxation is complete, directly reflecting the transient nature of dynamic heterogeneities. This also means that $\xi_{\mathrm{d}}(t)$ carries distinct signatures in different dynamical regimes. The choice of which dynamical window to probe it in is itself a physically meaningful decision. The prevailing picture, built almost exclusively on measurements at the $\alpha$-relaxation timescale, is that $\xi_{\mathrm{d}}(\tau_{\alpha})$ grows monotonically upon cooling towards $T_c$, yet remains finite and relatively modest in magnitude \cite{karmakar2009growing, flenner2015largeNat, zhang2018spatially, tah2020signature, li2020anatomy}.

Nevertheless, qualitative changes in the morphology of dynamic heterogeneities around $T_c$ have been reported in simulations \cite{stevenson2006shapes, royall2015strong, scalliet2022thirty, das2022crossover, herrero2024direct, pihlajamaa2025polydispersity} and experiments \cite{gokhale2016localized,ortlieb2023probing}, providing evidence that the evolution of dynamic correlations in this regime is non-trivial. Intriguingly, non-monotonic behavior of $\xi_{\mathrm{d}}(\tau_{\alpha})$ around $T_c$ has also been reported, but only in systems subject to external constraints such as partial pinning \cite{kob2012non, hima2015direct} or free surfaces \cite{peng2022nonmonotonic}. \citet{das2025collective} recently argued that the observations in partially pinned systems were a consequence of the pinning constraints rather than an intrinsic bulk property. All attempts to recover this non-monotonicity in a bulk, unconstrained system have so far been unsuccessful \cite{flenner2011analysis, flenner2012characterizing}. Given that these studies have without exception targeted $\xi_\mathrm{d}(\tau_{\alpha})$ in the $\alpha$-regime of relaxation, the possibility that the relevant signatures of the avoided transition have been sought in the wrong dynamical window has never been seriously explored.

A crucial recognition here comes from recalling that structural relaxation in supercooled liquids proceeds along two successive dynamical regimes: at intermediate times, the $\beta$-regime prevails, associated with fluctuations of particles about their surrounding `cages'. Only in the later $\alpha$-relaxation regime does long-range particle motion occur. In this sense, the $\beta$-relaxation regime is a crucial precursor to full structural relaxation. Caging dynamics necessarily shape and constrain the subsequent $\alpha$-process in the supercooled regime, especially near $T_c$ \cite{gao2025unified, caporaletti2026crossover} ; and yet it has been almost entirely overlooked in the context of dynamic correlation lengths.

This gap is all the more surprising given its privileged status in the theoretical landscape. The aforementioned mean-field theories of the dynamical glass transition \cite{gotze2009complex, maimbourg2016solution,parisi2020theory} all place the critical scenario in the $\beta$-regime. In those frameworks, $\tau_{\alpha}$ has no independent dynamical origin but is enslaved to a divergent $\beta$-timescale $\tau_{\beta}\propto(T-T_c)^{-1/2a}$ through $\tau_{\alpha}\propto (\tau_{\beta})^{1+a/b}$ with $a,b$ some material dependent exponents \cite{gotze1989beta}. Given the importance of this dynamical window, we argue that the eventual breakdown of the mean-field transition at $T_c$ should be most clearly imprinted on $\xi_{\mathrm{d}}(t)$ within the $\beta$-regime. To the best of our knowledge, only \citet{karmakar2016short} and \citet{tah2020signature} studied the dynamic correlation length in a bulk supercooled liquid in the $\beta$-regime. However, the temperature ranges explored did not extend below $T_c$, leaving the behavior of $\xi_{\mathrm{d}}(\tau_{\beta})$ across and below the crossover entirely uncharted.

Here, guided by the empirical and theoretical perspective above, we chart the behaviour of $\xi_{\mathrm{d}}(t)$ in the $\beta$-regime across $T_c$, capitalizing on recent computational advances that enable full equilibration of bulk glass-forming liquids well below the crossover \cite{berthier2023modern}. Our results demonstrate that $\xi_{\mathrm{d}}(t)$ in the $\beta$-regime exhibits non-monotonic temperature dependence across $T_c$. The picture our results suggest is one in which supercooled liquid dynamics become maximally cooperative at the crossover temperature $T_c$ and increasingly local at lower temperatures. Strikingly, this non-monotonic behavior finds a complete rationalization within the stochastic beta-relaxation theory \cite{rizzo2014long, rizzo2015qualitative, rizzo2016dynamical}, an extension of MCT beyond mean-field that explicitly predicts the scenario of an avoided dynamical transition at $T_c$ in physical dimensions. Taken together, our results provide direct and quantitative spatial evidence in favor of the avoided dynamical transition scenario near $T_c$, a scenario long anticipated theoretically.

\section{Results}
\begin{figure}
    \centering
    \includegraphics[width=\linewidth]{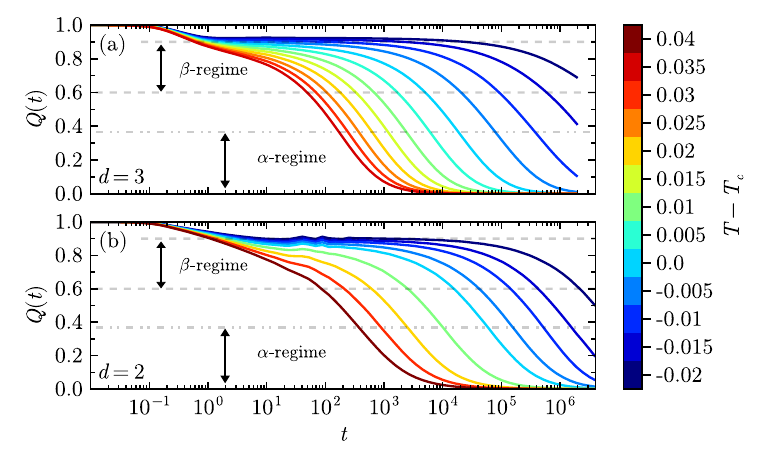}
    \caption{
    Two-step relaxation dynamics across the mode-coupling crossover. Overlap function $Q(t)$ at temperatures ranging from above to below $T_c$ (left to right: decreasing temperature) for the three- (panel (a), $N = 100\,000$ particles) and two-dimensional (panel (b), $N = 10\,000$ particles) model glass formers. The characteristic two-step decay develops upon cooling, with an intermediate plateau, the $\beta$-regime, becoming increasingly long-lived as $T_c$ is approached. Arrows indicate the approximate boundaries of the $\beta$- and $\alpha$-relaxation regimes as defined in the text. Temperatures are spaced as $(T -T_c) \in [-0.02, -0.015, \ldots, 0.035]$ in (a) and $(T-T_c) \in [-0.02, -0.015, \ldots, 0.0, 0.01, \ldots, 0.04]$ in (b). 
    }
    \label{fig:Overlap_Relax_2D_3D}
\end{figure}
\subsection{Characterization of Global Relaxation} 
We perform state-of-the-art equilibrium simulations of an established model glass-forming system in both two and three dimensions at and below the crossover $T_c$. Equilibrium configurations are generated using enhanced sampling techniques, from which molecular dynamics simulations are performed to obtain the particle trajectories. Simulation details are provided in the Supplementary Information (SI). For the model considered in this work, literature estimates give $T_c = 0.095$ (3D) and $T_c = 0.12$ (2D) \cite{scalliet2022thirty}.

\begin{figure*}[ht]
    \centering
\includegraphics[width=\linewidth]{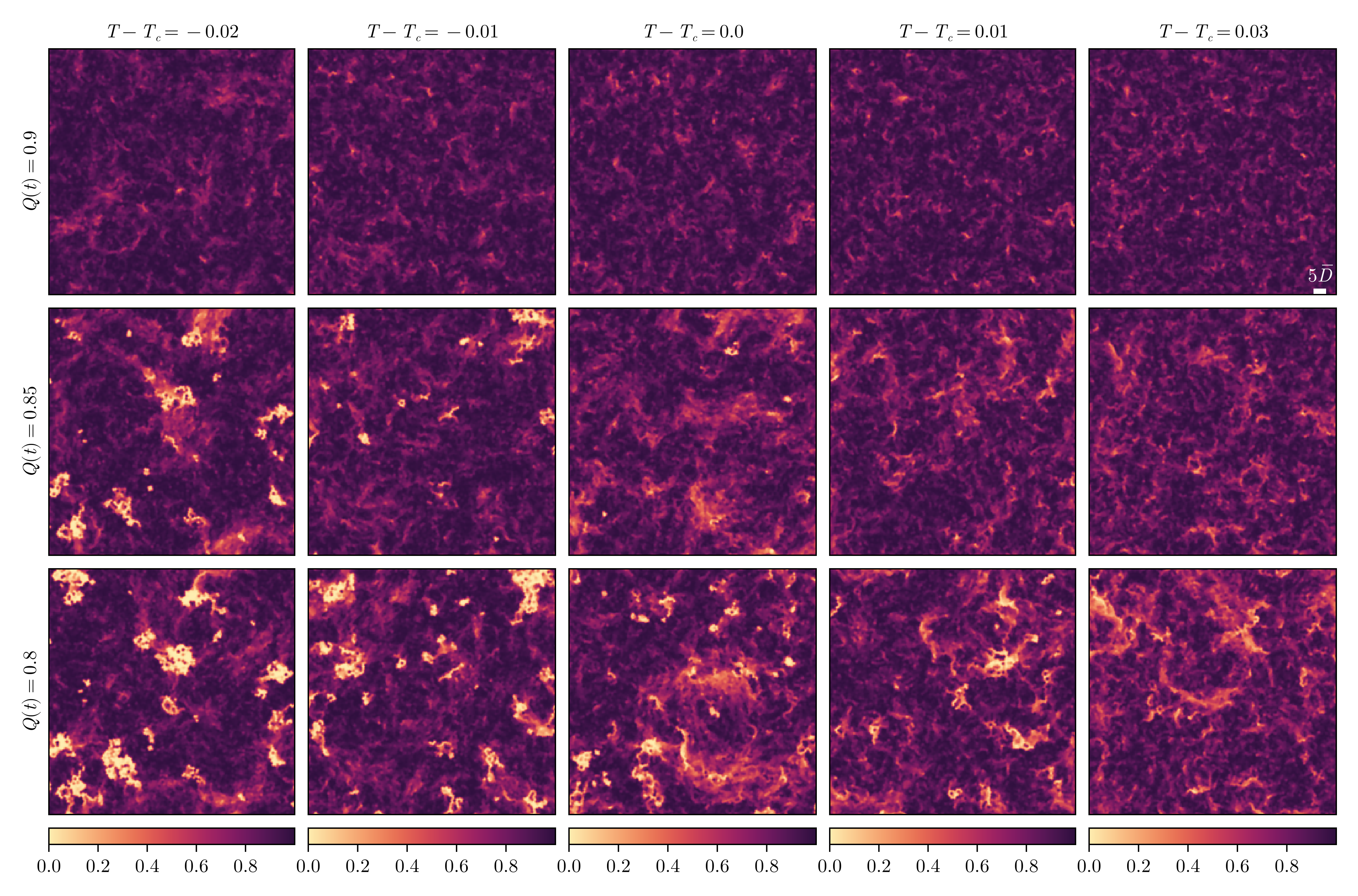}
    \caption{Spatial maps of the coarse-grained mobility field $\mu(\boldsymbol{r}, t)$ shown for the two-dimensional system considered in this work. Each column corresponds to a different temperature, while each row represents a distinct intrinsic time. Reading vertically reveals the temporal evolution of a single simulation at fixed temperature. Note that the mobility field is defined such that $\mu(\boldsymbol{r}, t) = 1.0$ indicates an immobile region and $\mu(\boldsymbol{r}, t) = 0.0$ a mobile one. A scale bar of size $5\overline{D}$ is shown in the top-right panel.}
    \label{fig:mobility_field_slab}
\end{figure*}

To quantify local relaxation in space and time, we use a particle-specific mobility indicator,
    \begin{equation}
        \mu_i(t) = \exp\left( -\frac{\Delta\boldsymbol{r}_i^2(t)}{2a^2} \right),
    \label{eq:mobility_indicator}
    \end{equation}
where $\Delta\boldsymbol{r}_i^2(t)$ is the squared displacement of particle $i$ over a time period $t$ and $a = 0.4$ is a coarse-graining length in units of mean particle diameter $\overline{D}$ \cite{dauchot2005dynamical}. We have verified that our results are qualitatively robust for different values of $a$ in the range $0.2$ to $0.8$, as well as for different definitions of the mobility indicator itself [e.g.\ the bond-breaking function, see SI].

The global relaxation of the system is monitored through the so-called overlap function
    \begin{equation}
        Q(t) = N^{-1}\langle \sum_{i=1}^N \mu_i(t) \rangle,
    \end{equation}
in which $\langle \cdot \rangle$ denotes an ensemble average. Exemplary results for $Q(t)$ at different temperatures around $T_c$ are shown in Fig.~\ref{fig:Overlap_Relax_2D_3D} for both three (a) and two (b) dimensions. The characteristic two-step decay of supercooled liquids is clearly visible: after an initial fast decay, $Q(t)$ exhibits an intermediate plateau, the $\beta$-regime, that becomes increasingly long-lived upon cooling, before the final decay to zero, the $\alpha$-regime, marks the completion of structural relaxation. Since relaxation times vary by orders of magnitude across the temperature range studied, we use the correlation function $Q(t)$ as a proxy for an internal clock of relaxation: reparameterizing time-dependent observables as functions of $Q(t)$ allows us to compare different systems at different temperatures at the same stage of relaxation. In the systems studied here, intrinsic time values $Q$ between 0.6 and 0.9 roughly correspond to the late and early $\beta$-relaxation regime respectively, while $Q \leq 1/e $ captures the $\alpha$-relaxation, as indicated in Fig.~\ref{fig:Overlap_Relax_2D_3D} (a).

While $Q(t)$ captures the global extent of structural relaxation, it is by construction blind to the spatial heterogeneity of local dynamics. To gain direct visual insight into the spatial structure of dynamic heterogeneities across $T_c$, we show in Fig.~\ref{fig:mobility_field_slab} snapshots of a locally coarse-grained mobility field
    \begin{equation}
        \mu(\boldsymbol{r}, t) = \frac{\sum_{i=1}^N \mu_i(t) w_i(\boldsymbol{r} ; a)}{ \sum_{i=1}^N w_i(\boldsymbol{r} ; a)},
    \end{equation}
with a Gaussian smoothing kernel $w_i(\boldsymbol{r} ; a) = e^{|\boldsymbol{r}-\boldsymbol{r}_i(t)|^2/2a^2}$ for a range of temperatures around $T_c$ (each shown in different columns) and intrinsic times within the $\beta$-regime (shown in rows). The smoothing factor $a$ of the kernel is the same as that of the mobility indicator ($a=0.4$). Examining the snapshots vertically, \textit{i.e.}~at fixed temperature reveals how the mobility field evolves dynamically. Above $T_c$ (right-most columns), the early indications of dynamic heterogeneity are visible. Upon cooling towards $T_c$, both the extent and magnitude of these heterogeneous regions increase. Below the crossover (left-most columns), the behavior changes qualitatively: mobile regions become noticeably more compact and the relaxation displays features reminiscent of dynamic facilitation \cite{scalliet2022thirty}. The compactification of mobile regions becomes even clearer when reading the figure horizontally, at fixed intrinsic time. As the temperature decreases (moving along a row from right to left), relaxation events first grow extended and with increasing intensity upon approaching $T_c$ from above and then become increasingly localized, yet even more intense, at lower temperatures. 

\subsection{Characterization of Fluctuations of Local Relaxation}
We quantify the dynamical fluctuations through the spatially resolved four-point correlation function $G_4(r,t)$  
    \begin{equation}
    G_4(r,t) = \frac{1}{N}\langle\sum_{i,j}\delta\mu_i(t)\delta\mu_j(t)
    \delta\left(\boldsymbol{r}-\boldsymbol{r}_{ij}(t)\right)\rangle,
    \label{eq:spatially_resolved_susceptibility}
    \end{equation}
where $\delta\mu_i(t)\equiv\mu_i(t)-Q(t)$ denotes the deviation of particle $i$'s mobility from the mean, and $\boldsymbol{r}_{ij}(t) = \boldsymbol{r}_i(t) - \boldsymbol{r}_j(t)$ is the instantaneous separation between particles $i$ and $j$. Equation \eqref{eq:spatially_resolved_susceptibility} thus measures how mobility fluctuations separated by distance $r$ at time $t$ are correlated \cite{berthier2011dynamical,berthier2011theoretical}.

In the mean-field picture of the dynamical glass transition, the emergence of the arrested phase is accompanied by a diverging susceptibility and correlation length \cite{biroli2006inhomogeneous, szamel2010diverging}. Translated to the four-point function Eq.~\eqref{eq:spatially_resolved_susceptibility}, this supposes a characteristic scaling form paralleling that of standard critical phenomena:
    \begin{equation}
        G_4(r,t) = \frac{A(t)}{r^{p}} f(r/\xi_{\mathrm{d}}(t)),
        \label{eq:G4_critical_scaling}
    \end{equation}
where $f(x) \sim \exp(-x)$ decays exponentially for sufficiently large $x$. Here, $p$ is some exponent and $A(t)$ a constant \cite{berthier2011theoretical}. The correlation length $\xi_{\mathrm{d}}(t)$ can then be determined from a direct fit of Eq.~\eqref{eq:G4_critical_scaling} or through the second-moment estimate 
	\begin{equation}
        \xi_{\mathrm{d}}(t)^2 \propto \frac{\int \mathrm{d}\boldsymbol{r} r^2 G_4(r,t)}{\int \mathrm{d}\boldsymbol{r} G_4(r,t)}
    \label{eq:lengthscale_operational_def}
    \end{equation}
close to $T_c$. These definitions are equivalent up to a multiplicative constant. One can also determine the correlation length $\xi_{\mathrm{d}}(t)$ through the low-$q$ behavior of the four-point dynamic structure factor, $S_4(q,t)$, which is the Fourier transform of $G_4(r,t)$ with respect to the separation variable $r$. The two approaches in principle yield the same correlation length. In practice, working in real space allows finite-size effects to be handled more directly (discussed further below); we therefore work exclusively on observables computed in real space.

If the crossover around $T_c$ is governed by a genuine phase transition, we expect $G_4(r,t)$ to exhibit critical scaling behavior, with a correlation length that diverges and an emergent scale invariance near $T_c$ \cite{biroli2006inhomogeneous, berthier2007spontaneous}. To investigate the relevance of the mean-field transition scenario, we obtain $G_4(r,t)$ and analyze its spatial profile in the vicinity of $T_c$. The key observation is that its growth does not follow the standard critical scaling form in full.

\begin{figure*}
    \centering
    \includegraphics[width=0.9\linewidth]{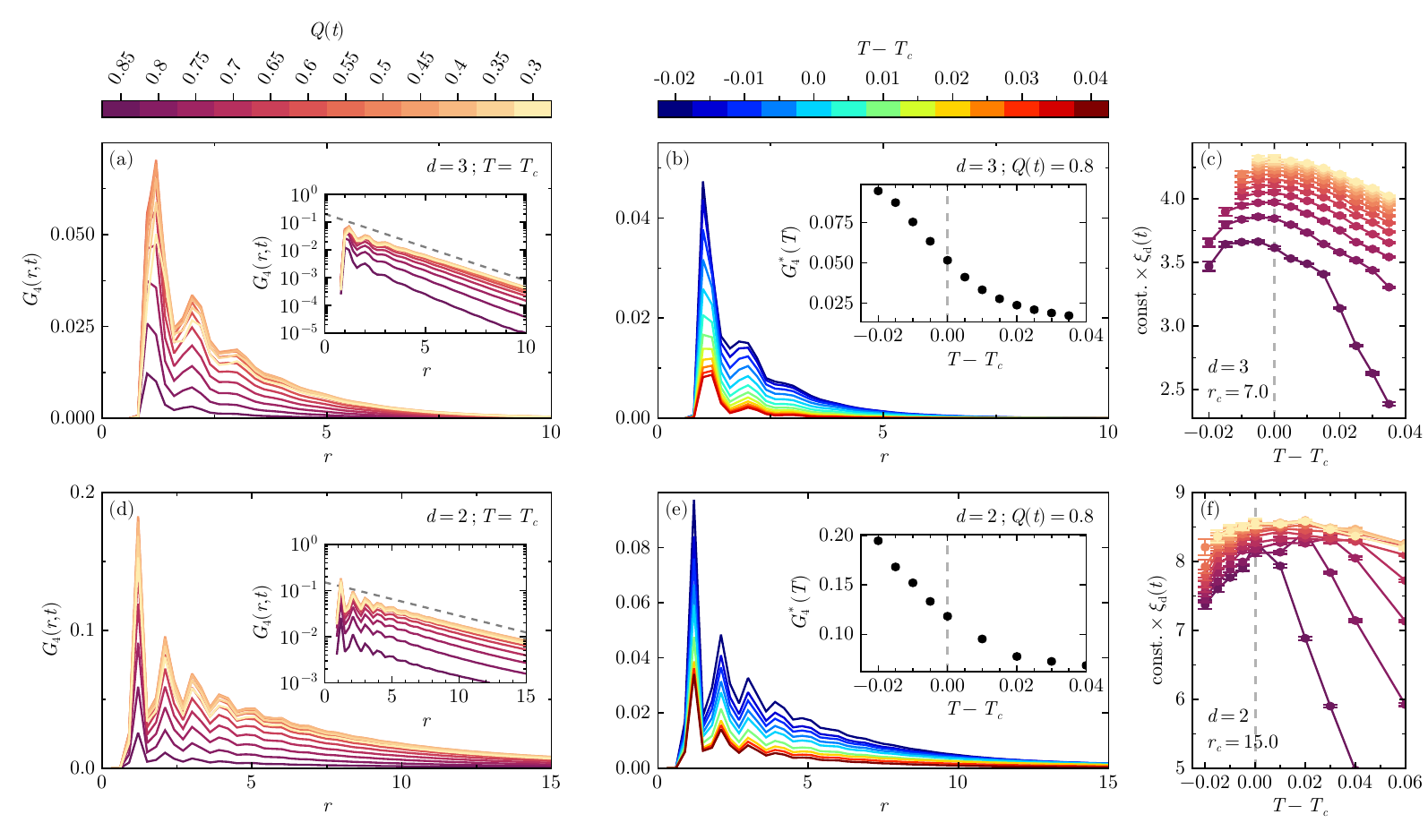}
    \caption{(a) Spatial profiles of $G_4(r,t)$ at fixed temperature $T = T_c$ for several intrinsic times in three dimensions ($N =$ 100 000). The inset shows the same data on a semi-logarithmic scale, with dashed lines an exponential decay serving as a guide to the eye. The curves correspond to all material times indicated in the colorbar: $Q(t) = 0.85, 0.08, \ldots, 0.3$. (b) Four-point correlation functions $G_4(r,t)$ at fixed intrinsic time $Q(t) = 0.8$ for varying temperatures $T$ relative to $T_c$ in three dimensions. The inset displays the corresponding maxima $G_4^*(T)$. The temperatures shown correspond to those shown in Fig.~\ref{fig:Overlap_Relax_2D_3D}(a). (c) Dynamical correlation length $\xi_{\mathrm{d}}(t)$, computed from the second-moment estimator Eq.~\eqref{eq:lengthscale_operational_def}, as a function of temperature in three dimensions. Error bars are obtained by propagating the standard error of the mean for each independent sample. The quantity $r_c$ shown in the panel defines the cut-off distance to which the integrals in Eq.~\eqref{eq:lengthscale_operational_def} were performed. See SI for a discussion of the cut-off dependence of our results. (d–f) Same quantities as in panels (a–c) but for the two-dimensional system ($N=10$ 000). Note that for panel (f), we have extended the $T-T_c$ axis to 0.6.
    }
\label{fig:xi_vs_Tc_sim_minimal}
\end{figure*}

We examine in Fig.~\ref{fig:xi_vs_Tc_sim_minimal}(a) the spatial profile of $G_4(r,t)$ at fixed $T=T_c$, in $d=3$, as structural relaxation progresses. During the early and intermediate stages of relaxation (the $\beta$-regime) we observe a pronounced growth of $G_4(r,t)$ at small $r$, with a clear exponential tail, shown in the semi-logarithmic inset. As the system enters the $\alpha$-relaxation regime $Q(t) \leq 1/e$, however, the magnitude of $G_4(r,t)$ begins to decrease. This behavior is expected: once long-range motion occurs, the mobility field must decorrelate and the associated measures of dynamic heterogeneities diminish. Dynamic heterogeneity is a transient phenomenon and $G_4(r,t)$ reflects this by decaying once the correlation function has dropped to $Q(t) \leq 1/e$. %This is also why the correlation length also decays at long times \cite{lacevic2002growing}.

In Fig.~\ref{fig:xi_vs_Tc_sim_minimal}(b), we turn to the spatial profile of $G_4(r,t)$ as a function of temperature, but at fixed intrinsic time, $Q(t) = 0.8$, in the $\beta$-regime. Again, the growth of $G_4(r,t)$ is concentrated at small $r$ upon cooling. The maximum of $G_4(r,t)$, denoted $G_4^*(T)$ at fixed intrinsic time (shown in the inset) continues to grow below $T_c$. Because local fluctuations of finite lifetime cannot grow without bound, this trend must however eventually saturate at sufficiently low temperatures \cite{montanari2006rigorous}. Crucially, $G_4(r,t)$ retains a clear exponential tail even at $T_c$, indicating that the associated $\xi_d(t)$ remains finite. We have verified that other values for $Q $and $T$ confirm this picture and thus show strong robustness of the results (see Figs.~SI.2-SI.3 in the SI).

Our observations reveal a clear but finite-ranged build-up of correlated motion approaching $T_c$. We next quantify the underlying spatial extent of these correlations through the dynamical correlation length $\xi_{\mathrm{d}}(t)$, determined via the second-moment estimator Eq.~\eqref{eq:lengthscale_operational_def}. To mitigate the presence of finite-size effects on $G_4(r,t)$, characterized by a negative tail at large distances, the radial integrals are truncated at a cut-off value $r_c$ that consistently exceeds $\xi_{\mathrm{d}}(t)$ by a factor 1.75--2, checked a posteriori, while remaining in a regime where the negative tails are sub-dominant (see SI for details). We have checked that variations in the precise value of $r_c$ do not affect the observations drawn below. As a further check, we have extracted $\xi_{\mathrm{d}}(t)$ through a direct exponential fitting of the tail of $G_4(r,t)$, restricting the fit to the range unaffected by finite-size effects; these results are consistent with the ones discussed below, and are presented in the SI. Both methods have been applied across multiple definitions of the mobility indicator, yielding several independent measurements that are in mutual qualitative agreement across all values of $Q$ in the $\beta$-regime.

The conclusions of this analysis are shown in Fig.~\ref{fig:xi_vs_Tc_sim_minimal}(c) for several intrinsic times across the fluid's structural relaxation, in three dimensions. We report that $\xi_{\mathrm{d}}(t)$ displays a non-monotonic dependence on temperature: it increases as the system approaches the crossover temperature $T_c$, reaches a maximum near $T_c$ and then decreases upon further cooling. This trend appears already in the early $\beta$-regime and persists through to the late $\beta$-regime. At the lowest temperatures investigated, our simulation time-window prevents us from accessing the full structural $\alpha$-relaxation regime in 3D and we cannot determine whether the non-monotonic behavior persists there. The monotonic increase that we observe in this regime is fully consistent with all previous studies of bulk glass formers. Figure~\ref{fig:xi_vs_Tc_sim_minimal}(d-f) shows the corresponding results for the two-dimensional system. The qualitative behavior parallels the three-dimensional case with only minor quantitative differences. In particular, oscillations at finite $r$ in $G_4(r,t)$ are more pronounced in two dimensions, reflecting stronger structural ordering in this model \cite{tong2023emerging}. Moreover, the overall magnitude of $G_4(r,t)$ is larger in $d=2$, consistent with the more strongly cooperative dynamics reported for two-dimensional glass formers \cite{pihlajamaa2025polydispersity}. Shown in panel (f), the non-monotonicity of $\xi_{\mathrm{d}}(t)$ in the $\beta$-regime is evident. Furthermore, the reduced computational cost of two-dimensional simulations allows us to resolve the $\alpha$-regime below the crossover temperature, where $\xi_{\mathrm{d}}(\tau_{\alpha})$ exhibits a clear saturation upon cooling, in agreement with earlier studies \cite{flenner2012characterizing}.

Our observations thus reveal a non-trivial evolution of dynamical correlations across the mode-coupling crossover: $\xi_{\mathrm{d}}(t)$ grows on approaching $T_c$ but decreases again upon further cooling, even as local fluctuations intensify. This behavior suggests an increasingly short-ranged and intermittent mobility below the crossover and motivates the need for a theoretical description that captures both the spatial growth and subsequent localization of correlated dynamical fluctuations. 

% \section*{Theoretical Account}
\subsection{Theoretical Description of $\xi_{\mathrm{d}}(t)$}
We propose an interpretation of these findings consistent with the idea of an avoided dynamical phase transition at $T_c$. Specifically, we use the framework of stochastic beta-relaxation theory (SBR): a beyond-mean-field theory that describes how finite-dimensional, long-wavelength critical fluctuations smear out the dynamical transition predicted by the mean-field, replacing it with a smooth dynamical crossover \cite{rizzo2014long, rizzo2015qualitative, rizzo2016dynamical}. 

The central idea of SBR is that in the vicinity of an avoided transition the correlation length, albeit finite, is sufficiently large that the system can be described by a coarse-grained effective theory. Remarkably this implies that the same effective theory can describe systems that are utterly different at the microscopic level, in direct analogy with the concept of universality near true phase transitions. Various arguments constrain the form of this effective theory \cite{rizzo2014long, rizzo2016dynamical}, and a non-trivial computation shows that this theory is equivalent to the SBR equations  described below. Consistent with universality, SBR has been shown to describe both spin-glasses and kinetically constrained models \cite{wolynes2023non,rizzo2020solvable}. Here we present the first application to numerical data for supercooled liquids.
%Physically, SBR captures the idea that the distance to the transition, $\sigma \propto (T_c-T)$, varies locally in space, so that structural relaxation proceeds through the isotropic growth of liquid-like regions into more glassy ones. 

At the quantitative level, SBR describes the deviation from the plateau of the correlation function, $Q(a;t)$, at fixed microscopic probing length $a$, which we explicitly re-introduce in our notation for the following discussion. Expanding around the plateau value yields
    \begin{equation}
        Q(a; t) = Q_p(a) + H(a) G(t),
    \end{equation}
where $Q_p(a)$ is the plateau height and the deviation is factorized into a static amplitude $H(a)$ and a time-dependent function $G(t)$ \cite{gotze1989beta}. The deviation $G(t)$ is expressed as the spatial average of a coarse-grained, time-dependent field $g(\boldsymbol{x}, t)$
    \begin{equation}
        G(t) = \frac{1}{V} \int \mathrm{d}\boldsymbol{x}\ g(\boldsymbol{x}, t) \ .
    \end{equation}
Specifically, $g(\boldsymbol{x}, t)$ represents the local degree of structural relaxation and should be interpreted analogously to the coarse grained mobility field $\mu(\boldsymbol{x}, t)$ introduced earlier. Within SBR, the field $g(\boldsymbol{x}, t)$ obeys the following stochastic integro-differential equation 
\begin{widetext}
    \begin{equation}
    % \begin{split}
    s(\boldsymbol{x}) = -\alpha \nabla^2 g(\boldsymbol{x},t)
    - \lambda g(\boldsymbol{x},t)^2 + \frac{\mathrm{d}}{\mathrm{d}t}\int_0^t \mathrm{d}\tau\,
    g(\boldsymbol{x},t-\tau)\,g(\boldsymbol{x},\tau),
    % \end{split}
    \label{eq:SBR}
    \end{equation}
\end{widetext}
where $\alpha$ and $\lambda$ are material-dependent coupling constants, and $s(\boldsymbol{x}) = \sigma + \delta s(\boldsymbol{x})$ is a local separation parameter consisting of a mean distance $\sigma$ from the critical temperature $T_c$ and quenched spatial fluctuations $\delta s(\boldsymbol{x})$ around it. The quenched disorder is Gaussian, with zero mean $\llbracket \delta s(\boldsymbol{x}) \rrbracket = 0$ and short-ranged correlations $\llbracket \delta s(\boldsymbol{x})\delta s(\boldsymbol{y})\rrbracket = \Delta s^2 \delta(\boldsymbol{x}-\boldsymbol{y})$, where $\llbracket \cdot \rrbracket$ denotes an average over the disorder. Physical observables are obtained by averaging over this disorder, which in the thermodynamic limit is equivalent to spatial self-averaging over the quenched field. 

\begin{figure}
    \centering
    \includegraphics[width=\linewidth]{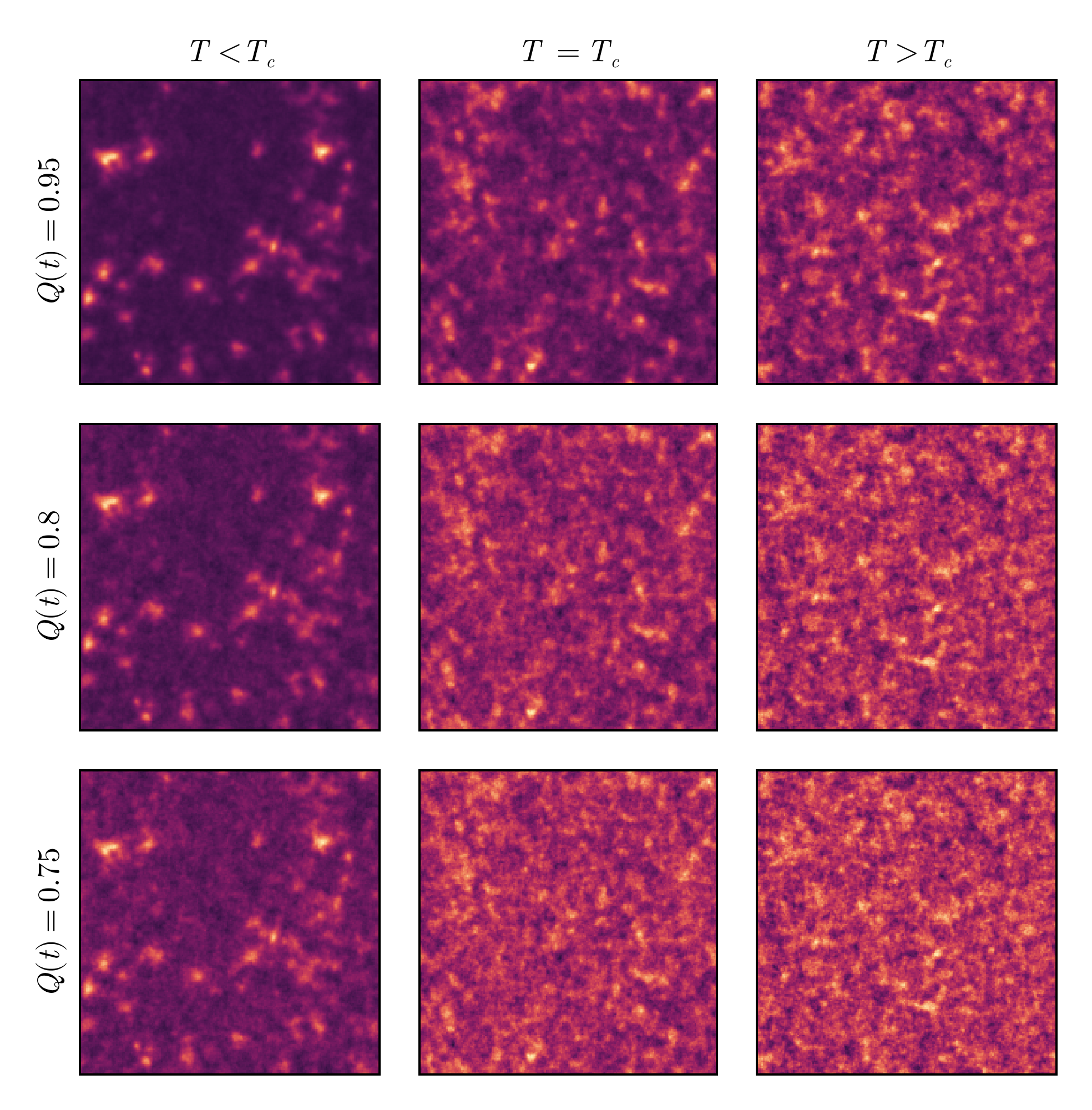}
    \caption{SBR solutions in two dimensions for several temperatures (columns) and intrinsic times (rows). The spatial features of the field $g(\boldsymbol{x},t)$ qualitatively mirror the simulated mobility maps, indicating that SBR captures the key dynamic heterogeneity in the $\beta$-regime near $T_c$. Note that the colorscheme is analogous to that used in Fig.~\ref{fig:mobility_field_slab}, where dark purple indicates immobility and light yellow indicates mobility.}
    \label{fig:beta_scaling_2D}
\end{figure}

SBR has a clear physical interpretation: the local order parameter $g(\boldsymbol{x}, t)$ relaxes in the presence of quenched local fluctuations of the temperature, so that even below $T_c$ there exist rare regions whose local temperature corresponds to the liquid phase, through which the mean-field transition is avoided. This picture naturally gives rise to dynamic heterogeneity: above $T_c$, the mean local temperature corresponds to the liquid phase $(\sigma < 0)$ and is not significantly affected by the quenched fluctuations, leading to homogeneous spatiotemporal fluctuations in the dynamics, as shown in the rightmost column of Fig.~\ref{fig:beta_scaling_2D}. By contrast, below $T_c$, the mean local temperature corresponds to the mean-field glass $(\sigma > 0)$. Relaxation is then driven exclusively by rare regions where fluctuations in the local separation parameter  are large enough to restore liquid-like behaviour {\it i.e.} $\sigma + \delta s(x)<0$, as seen in the leftmost column of Fig.~\ref{fig:beta_scaling_2D}. The observation of spatiotemporally heterogeneous dynamics is a direct consequence of isolated liquid-like regions that are rare because for large positive $\sigma$ the local fluctuation $\delta s(x)$ must be in the negative tail of the Gaussian in order to make $\sigma + \delta s(x)<0$. The representative solutions shown in Fig.~\ref{fig:beta_scaling_2D} closely resemble the mobility maps obtained in our simulations [c.f.~Fig.~\ref{fig:mobility_field_slab}], providing a first qualitative confirmation that SBR captures the essential spatiotemporal features of structural relaxation in the $\beta$-regime around $T_c$. Numerical details concerning the solution of Eq.~\eqref{eq:SBR} are provided in the SI. 

Because SBR describes a translationally invariant system, we consider the spatial correlations of the local relaxation field through
\begin{equation}
    \begin{split}
        G_4^{\mathrm{SBR}}(r, t) =&\ \frac{1}{V} \int \mathrm{d}\boldsymbol{x} g(\boldsymbol{x}+\boldsymbol{r}, t) g(\boldsymbol{x}, t)-G(t)^2.
    \end{split}
    \label{eq:G4_SBR}
    \end{equation}
This four-point function plays the same role as its simulation counterpart Eq.~\eqref{eq:spatially_resolved_susceptibility}, and allows us to define a dynamical correlation length $\xi_{\mathrm{d}}^{\mathrm{SBR}}(t)$ using the same second-moment estimator Eq.~\eqref{eq:lengthscale_operational_def}. Finite-size effects on $G_4^{\mathrm{SBR}}(r, t)$ from the finite grid size are handled in the same way as in the simulations. 

A quantitative comparison between SBR and simulation requires fixing the theory's parameters $\alpha, \lambda, \sigma$ and $\Delta s^2$. In principle these can be derived microscopically \cite{franz2012quantitative, parisi2013critical, laudicina2025theory}, but in the present context they must be fitted. This fitting is non-trivial for two reasons. First, the crossover regime around $T_c$ is a region where multiple physical mechanisms, not necessarily contained in SBR, start to compete \cite{scalliet2019nature, herrero2024direct}. Second, the simulation signal carries contributions from the underlying liquid structure, which are not captured by the continuum SBR description and further hamper the comparison. A fully consistent simultaneous fit of $G(t)$, $G_4(r,t)$ and $\xi_{\mathrm{d}}(t)$ across all temperatures and timescales would require a more systematic parameter estimation procedure that lies beyond the scope of the present work. Here, we therefore fit the parameters to best reproduce the central observable of interest $\xi_{\mathrm{d}}(t)$ alone, and leave a more comprehensive comparison for future work. The fitting method is detailed in the SI, and the numericla values are shown in Table~\ref{tab:SBR_params}.

\begin{table}
\centering
\begin{tabular}{||l|c|c||}
\hline
Spatial Dimension & $d=3$ & $d=2$ \\
\hline
Critical Temperature $T_c$ & $0.117$ & $0.130$ \\
\hline
Plateau Value $Q_p$ & $0.856$ & $0.800$ \\
\hline
Exponent Parameter $\lambda$ & $0.74$ & $0.78$ \\
\hline
Reduced Temp. prefactor $\sigma/(T_c - T)$ & $-0.0328$ & $-0.125$ \\
\hline
Stiffness $\alpha$ & $0.373$ & $0.0562$ \\
\hline
Disorder Strength $\Delta \sigma$ & $0.0391$ & $0.00832$ \\
\hline
Grid Spacing $\Delta x$ & $2.4$ & $0.375$ \\
\hline
\end{tabular}
\caption{SBR parameters used to generate the theoretical results in 
Fig.~\ref{fig:xi_vs_Tc_SBR} of the main text. The exponent parameter $\lambda$ is taken from \citet{scalliet2022thirty} for the same model system; all other parameters are fitted to best reproduce $\xi_{\mathrm{d}}(t)$.}
\label{tab:SBR_params}
\end{table}

\begin{figure}[ht]
    \centering
    \includegraphics[width=\linewidth]{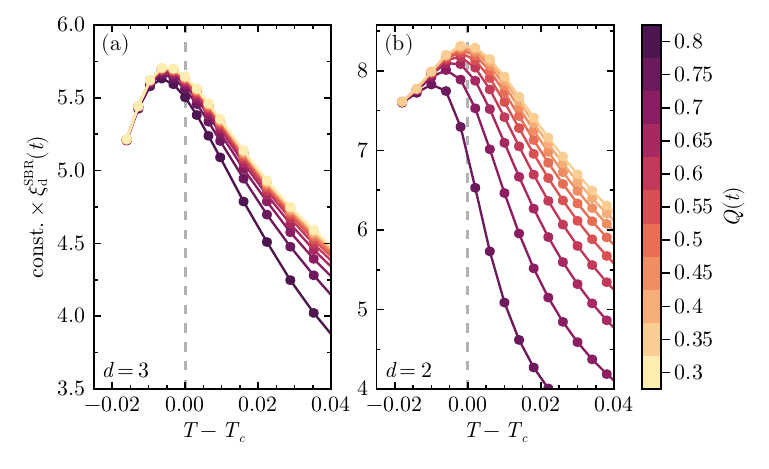}
    \caption{SBR predictions for the dynamical correlation length $\xi_{\mathrm{d}}^{\mathrm{SBR}}(t)$ in (a) three and (b) two dimensions, obtained from the second-moment estimator Eq.~\eqref{eq:lengthscale_operational_def} applied to $G_4^{\mathrm{SBR}}(r,t)$ defined in Eq.~\eqref{eq:G4_SBR}. Results are shown for intrinsic times $Q \in [0.8, 0.75, \ldots, 0.3]$ in (a) and $Q \in [0.8, 0.75, \ldots, 0.4]$ in (b). The non-monotonic temperature dependence of $\xi_{\mathrm{d}}^{\mathrm{SBR}}(t)$ across $T_c$ is reproduced across all intrinsic times and in both dimensions.
}
    \label{fig:xi_vs_Tc_SBR}
\end{figure}

In Fig.~\ref{fig:xi_vs_Tc_SBR}, we show the dynamic correlation length $\xi_{\mathrm{d}}^{\mathrm{SBR}}(t)$ predicted by the SBR equations in three dimensions [panel (a)] and in two dimensions [panel (b)]. In both panels, the results are plotted against the temperature difference to the critical point and for different intrinsic times. Across both dimensions and throughout the entire $\beta$-regime, SBR predicts a clear non-monotonic dependence of the correlation length within the $\beta$-regime, in excellent qualitative and near-quantitative agreement with the simulations. Note that in SBR, an overall unit of length remains undetermined in comparing the value of $\xi_\text{d}^\text{SBR}$ to the simulation $\xi_\text{d}$. A noteworthy feature of the SBR prediction is that the maximum of the correlation length occurs just below the mean-field critical temperature. This feature is most clearly observed in 3D, and slightly less apparent in the two-dimensional data, where the position of the maximum is more sensitive to the choice of $Q(t)$. This precisely mirrors the trend observed in the simulation data [c.f. Fig.\ref{fig:xi_vs_Tc_sim_minimal}(c) and (f)], though we mention that the estimation of $T_c$ there carries systematic errors that cannot be quantified. SBR also reproduces that the maximum of $\xi_\text{d}(t)$ is higher in the 2D system than it is in 3D (by a factor of $1.5$, compared to about $2$ in the simulation). We stress that the non-monotonic behavior of the correlation length is a generic qualitative feature of SBR (see discussion in Ref.~\cite{rizzo2014long}) for \emph{all} choices of the quantitative parameters. We are not aware of any other theory that yields a comprehensive dynamical description leading to the behavior of Fig.~\ref{fig:xi_vs_Tc_SBR}.

At low intrinsic time (corresponding to the onset of the $\alpha$-regime), the SBR correlation length $\xi_{\mathrm{d}}^{\mathrm{SBR}}(t)$ as a function of the temperature exhibits a clear saturation to a limiting curve. This occurs because the solution of the SBR solutions take the form $g(\boldsymbol{x},t) \approx -\, B(\boldsymbol{x})t^b$ at large values of $t$, with $B(\boldsymbol{x})$ depending in a complex way on the quenched temperature shifts $s(\boldsymbol{x})$, and on temperature. The curves of Fig.~\ref{fig:xi_vs_Tc_SBR} saturate essentially to the correlation length of the $B(\boldsymbol{x})$ field as a function of the temperature \cite{rizzo2015nature}. A close inspection of Fig.~\ref{fig:xi_vs_Tc_sim_minimal}-(c) and (f) suggests that this saturation is not exhibited by the simulation data in the $\alpha$-regime. At fixed temperature, the correlation length appears to keep growing with decreasing intrinsic time, though slower than in the $\beta$-regime. This is not unexpected: SBR is an effective theory of dynamical fluctuations in the $\beta$-regime, and extending its quantitative predictions to the $\alpha$-regime lies beyond its current scope.

\subsection{Magnitude of Dynamical Fluctuations} 

The results discussed above raise the question of whether the decreasing size of cooperative regions below $T_c$ is accompanied by a weakening of dynamic heterogeneities. To address this, we examine the four-point susceptibility
    \begin{equation}
        \chi_4(t)= \frac{1}{V} \int \mathrm{d}\boldsymbol{r}\, G_4(r,t)
    \end{equation}
which measures the total amplitude of correlated motion in our simulations. We compute $\chi_4(t)$ by integrating the full spatial profile of $G_4(r,t)$ supplemented by fitted exponential tails to correct for finite-size effects, rather than through the direct estimator $N^{-1}\langle \sum_{i,j}\delta\mu_i(t)\delta\mu_j(t)\rangle$, which carries well-known system-size and ensemble dependencies requiring careful treatment \cite{berthier2007spontaneous}.
\begin{figure}
    \centering
    \includegraphics[width=\linewidth]{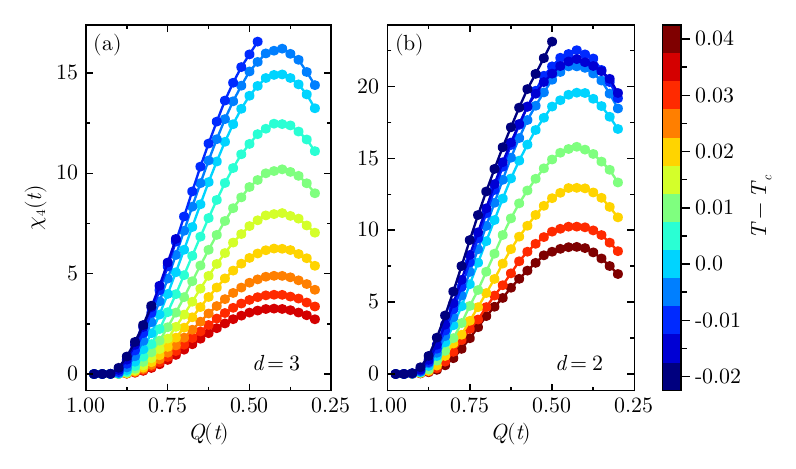}
    \caption{Dynamical susceptibility $\chi_4(t)$ for the three- (a) and two-dimensional (b) systems, evaluated at temperatures around the mode-coupling crossover $T_c$.}
    \label{fig:chi4_vs_Qt_2D_3D}
\end{figure}

The resulting susceptibilities are shown in Fig.~\ref{fig:chi4_vs_Qt_2D_3D} for three (a) and two (b) dimensions respectively. In agreement with an extensive body of simulation and experimental literature \cite{berthier2004time, szamel2006four, chandler2006lengthscale, berthier2011dynamical}, $\chi_4(t)$ grows markedly upon cooling and displays a clear peak near the structural relaxation time. In the $\beta$-regime, where our measurements of $\xi_{\mathrm{d}}(t)$ are most reliable, $\chi_4(t)$ grows monotonically throughout. This demonstrates that, contrary to conventional interpretations, an increasing dynamical susceptibility does not necessarily imply a corresponding increase in the spatial extent of correlated fluctuations. Spatial correlation lengths and global susceptibilities encode complementary but not interchangeable information about glassy dynamics.

\section{Discussion}
Using computer simulations of a model supercooled liquid in two and three dimensions, we have shown that the $\beta$-regime dynamical correlation length $\xi_{\mathrm{d}}(t)$ exhibits a non-monotonic temperature dependence across the mode-coupling crossover: it grows upon cooling, reaches a maximum near $T_c$, and decreases upon further cooling. The peak reflects a regime of maximal cooperative motion, with  correlated regions reaching up to 5 particle diameters in three dimensions, followed by a crossover to increasingly compact and localized relaxation events below $T_c$. We note that this modest magnitude is in fact fully commensurate with the predictions of mean-field theory: the divergence of $\xi_{\mathrm{d}}$ at $T_c$ is algebraically slow, $\xi_{\mathrm{d}} \propto |T-T_c|^{-1/4}$, and explicit mode-coupling calculations for a hard-sphere systems yield correlation lengths of only a few particle diameters even extremely close to the critical point \cite{laudicina2025theory}. Our findings therefore lend direct and quantitative support to the avoided-transition interpretation of the mode-coupling crossover, and in particular to the Stochastic Beta-Relaxation theory \cite{rizzo2014long, rizzo2015qualitative, rizzo2016dynamical}, an extension of MCT beyond mean-field that explicitly predicts a non-monotonic dynamical correlation length in the $\beta$-regime as a hallmark of the avoided phase transition.

Our results are fully consistent with the prevailing view of a monotonically growing dynamical length scale in the $\alpha$-regime: the non-monotonicity we report is a feature of the $\beta$-regime only. Our observations are therefore complementary rather than contradictory to existing literature, and together paint a richer picture of how dynamic correlations evolve around the crossover. Additionally, our results establish an operational, parameter-free definition of $T_c$ as the temperature at which the $\beta$-regime dynamical correlation length is maximal. This provides an alternative to the conventional determination of $T_c$ from power-law fits of transport coefficients, which carry methodological uncertainties.

We stress that all conclusions drawn here pertain specifically to the dynamics near $T_c$ and are also fully compatible with the existence of distinct physics at lower temperatures. The question of how the spatially extended, maximally cooperative dynamics near $T_c$ gives way to the so-called `activated processes' that govern the deeply supercooled regime is a fascinating one. In particular, the localization of dynamic heterogeneities below $T_c$ that we report may itself be an early signature of this crossover to activation-dominated dynamics. Whether this evolution connects continuously to the mosaic-driven processes advocated by theories of the deeply supercooled regime, and how the two lengthscales governing these respective regimes relate to one another \cite{biroli2012random}, are questions that we hope the present work will motivate.

Looking ahead, it would also be valuable to test the generality of our results extending the present analysis to other model systems such as the Kob-Andersen binary mixture recently studied below $T_c$ \citet{rusciano2025low}, and to high-dimensional hard-sphere systems where non-perturbative corrections to mean-field are expected to be reduced and the pseudo-critical window around $T_c$ broader and more sharply delineated \cite{charbonneau2022dimensional}. From an experimental standpoint, the $\beta$-regime is in principle accessible in dense colloidal suspensions and granular systems, where four-point dynamic correlations have already been measured at the $\alpha$-timescale \cite{zhang2011cooperative}. We hope the present results will motivate efforts in this direction, so as to provide direct experimental confirmation of the avoided MCT transition scenario presented here.

\textbf{Acknowledgements}: CCLL, IP and LMCJ acknowledge funding from a Vidi grant from the Dutch Research Council. 
TR acknowledges support from the 2021 first FIS (Fondo Italiano per la Scienza) funding scheme (FIS783 - SMaC - Statistical Mechanics and Complexity University and Research) from Italian MUR (Ministry of University and Research).
We gratefully thank G. Szamel for insightful discussions and comments on this work.

\bibliography{bibliography}

\pagebreak
\onecolumngrid

\begin{center}
\textbf{\large Supplementary Information: Non-Monotonic Dynamical Correlations Across The Glass Crossover}
\end{center}
%%%%%%%%%% Merge with supplemental materials %%%%%%%%%%
%%%%%%%%%% Prefix a "S" to all equations, figures, tables and reset the counter %%%%%%%%%%
\setcounter{equation}{0}
\setcounter{figure}{0}
\setcounter{table}{0}
\setcounter{section}{0}
\setcounter{page}{1}
\renewcommand{\thesection}{S\arabic{section}}
\renewcommand{\theequation}{S\arabic{equation}}
\renewcommand{\thefigure}{S\arabic{figure}}
\renewcommand{\thetable}{S\arabic{table}}
\renewcommand{\bibnumfmt}[1]{[S#1]}
\renewcommand{\citenumfont}[1]{S#1}

This Supplementary Information presents additional results that supplement the analysis discussed in the main text, as well as further details that pertain to the numerical details of the work.

\setcounter{figure}{0}
\renewcommand{\thefigure}{SI.\arabic{figure}}

\section{Model System \& Simulation Protocols} \label{app:simulation_protocol}

We investigate a recently proposed polydisperse mixture introduced in Refs.~\cite{berthier2023modern, scalliet2022thirty}. This system consists of $N$ particles interacting via the pair potential
    \begin{equation}
        u(r_{ij}) = \epsilon \left(\frac{r_{ij}}{D_{ij}}\right)^{-12} + c_4\left(\frac{r_{ij}}{D_{ij}}\right)^4 + c_2\left(\frac{r_{ij}}{D_{ij}}\right)^2 + c_0,
    \end{equation}
for $r_{ij}/D_{ij} < r_c$, with $u(r_{ij}) = 0$ otherwise. The constants $c_0 = -28\epsilon/r_c^{12}$, $c_2 = 48\epsilon/r_c^{14}$ and $c_4 = -21\epsilon/r_c^{16}$ are chosen to ensure that the potential is both continuous and twice differentiable at the cutoff radius $r_c$. The pair diameter is defined as
    \begin{equation}
        D_{ij} = \frac{1}{2}(D_i + D_j)(1 - \zeta |D_i - D_j|),
    \end{equation}
where the nonadditivity parameter $\zeta = 0.2$ suppresses crystallisation by promoting demixing. The individual particle diameters are drawn from a polydispersity distribution
    \begin{equation}
        P(D) = \frac{A}{D^3}, \quad D \in [D_{\mathrm{min}}, D_{\mathrm{max}}],
    \end{equation}
where the normalisation factor is given by $A = D_{\mathrm{min}}D_{\mathrm{max}} / (D_{\mathrm{max}} - D_{\mathrm{min}})$ and the lower cutoff is set as $D_{\mathrm{min}} = D_{\mathrm{max}} / (2 D_{\mathrm{max}} - 1)$. These choices ensure that the mean diameter, $\overline{D}$, is unity and that the distribution is properly normalised. The maximal size ratio is fixed at $D_{\mathrm{large}}/D_{\mathrm{small}} = 2.219$ in both two- and three-dimensional systems.

To generate well-equilibrated initial configurations, we employ the SWAP Monte Carlo (SMC) algorithm for systems of size $N =$ 1 000, 10 000 and 100 000 in spatial dimension $d=3$ and $N=$ 1 000 and 10 000 in $d=2$. All simulates are performed at a fixed number density $n=1$. For system sizes of $N=$ 1 000 in both $d=2,3$, for each temperature $T$, we generate at least 100 independent equilibrium configurations, ensuring equilibration by running the SMC algorithm for at least $100 \tau^{\mathrm{SMC}}_\alpha$, where $\tau^{\mathrm{SMC}}_\alpha$ is the $\alpha$-relaxation time of the SMC dynamics. This timescale is defined as the time at which the overlap function $Q(t)$ has decayed to $1/e$. For system sizes of $N =$ 10 000, 100 000 particles in both dimensions investigated, we generate 10 independent equilibrium configurations ensuring equilibration by running the SMC algorithm for at least $6 \tau^{\mathrm{SMC}}_\alpha$. 

Each equilibrated configuration is then used as the initial state for a molecular dynamics (MD) simulation performed in the microcanonical (NVE) ensemble. The simulations use a time step of $\Delta t = 0.01$ in reduced units, where time is expressed in terms of $(m \overline{D}^2 / \epsilon)^{1/2}$. To ensure adequate sampling of long-time dynamics, we run up to at least $10^8$ MD steps at the lowest investigated temperature, $T = 0.07$ (3D) and $T=0.08$ (2D). For this model, literature estimates give $T_c = 0.095$ (3D) and $T_c = 0.12$ (2D) \cite{scalliet2022thirty}. This places the temperatures investigated substantially below the crossover. 

\section{Additional Numerical Details}\label{app:additional_results_lengthscale}
\setcounter{figure}{0}

This section provides additional information regarding the definition and computation of the various observables presented in the main text. 

\subsection{The Self-Overlap Function}\label{app:self_overlap}

We begin by characterizing the global dynamical behavior of the system at the dynamical crossover. Specifically, we consider the time-dependent correlation function 
\begin{equation}
    Q(t) = \frac{1}{N}\langle[\sum_{i=1}^N \mu_i(t) ]\rangle,
    \label{eq:correlation_function}
\end{equation}
where where $\mu_i(t)$ represents a particle-specific mobility indicator for particle $i$ and the angular brackets $\langle ... \rangle$ denote a statistical average over the statistical ensemble and $[...]$ is a time average over a single run. Since we work at equilibrium, this distinction was not made in the main text as they are equivalent. In this study, we define the single-particle mobility using 
\begin{equation}
    \mu_i(t) = \exp\left( -\frac{\Delta r_i^2(t)}{2a^2} \right),
\end{equation} 
with $\Delta r_i^2(t)$ denoting the squared displacement of particle $i$ at time $t$ and $a$ serving as a coarse-graining factor. The coarse-graining factor $a$ (set to $a=0.4$ in this work) is introduced to suppress short-time cage rattling in the particle displacements, ensuring that the mobility indicator captures genuine particle rearrangements. We have checked that all results presented here are insensitive to the precise choice of $a$ or the mobility indicator itself, which ultimately only amount to minor quantitative differences in the results.
\begin{figure*}[ht]
    \centering
    \includegraphics[width=\linewidth]{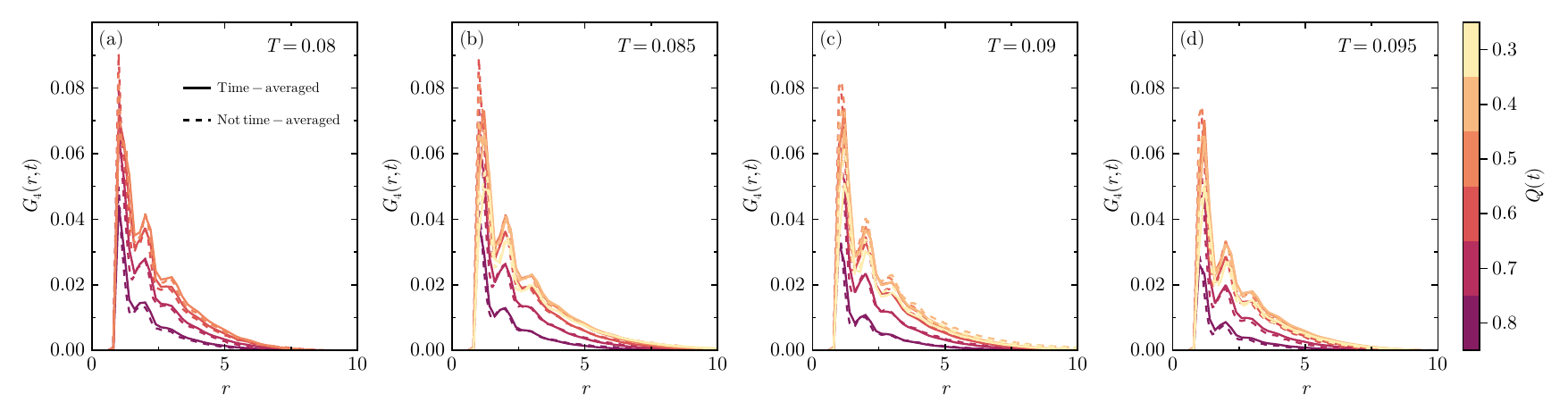}
    \caption{Comparison of the spatial profile of $G_4(r,t)$ for different intrinsic times and different temperatures below the mode-coupling crossover for a system comprised of $N = 10$ 000 particles in 3D for systems in which a time-average was or was not (dashed lines) performed. Legend is the same as Fig.~3 (main text).}    \label{fig:comparison_G4_time_averaging_3D_N_10000} 
\end{figure*}
In practice, the ``estimator" for $Q(t)$ is obtained by considering
% \begin{widetext}
    \begin{equation}
        Q(t) \approx \left( \frac{1}{N_{\mathrm{samples}}} \sum_{n=1}^{N_{\mathrm{samples}}} \left( \frac{1}{N_{\mathrm{pairs}}} \sum_{i_p=1}^{N_{\mathrm{pairs}}} \left(\frac{1}{N}\sum_{i=1}^N\mu_i^{(n)}(\Delta t_{i_p})\right)\right)\right)
    \label{eq:self_overlap_estimator}
    \end{equation}
% \end{widetext}
where $\mu_i^{(n)}(\Delta t_{i_p})$ is the mobility of particle $i$ in simulation $n$ in time interval $\Delta t_{i_p}$ of size $t$. Here, $N_{\mathrm{pairs}}$ corresponds to the number of time-intervals of size $t$ in a given run. We therefore perform an average over different time-origins in a given run and then an average over different realizations of the system. Here, $N_{\mathrm{sample}}$ denotes the number of independent simulation runs. We next describe how the time-averaging is performed in a given run. Specifically, we divide the full trajectory of $N_{\mathrm{MD}}$ MD integration steps into $N_{\mathrm{steps}}$ evenly spaced starting offsets. From each offset, we record configurations at times that grow geometrically according to a factor of $1.3$. We take $N_{\mathrm{steps}} = 100$ in our simulations and $N_{\mathrm{MD}} \approx 10^8-10^9$ depending on system size. When indicated, the error bars correspond to the standard error of the mean for the $N_{\mathrm{samples}}$ independent samples. (We thus treat the time-averaged quantities as if they were one measurement).
\begin{figure*}
    \centering
    \includegraphics[width=\linewidth]{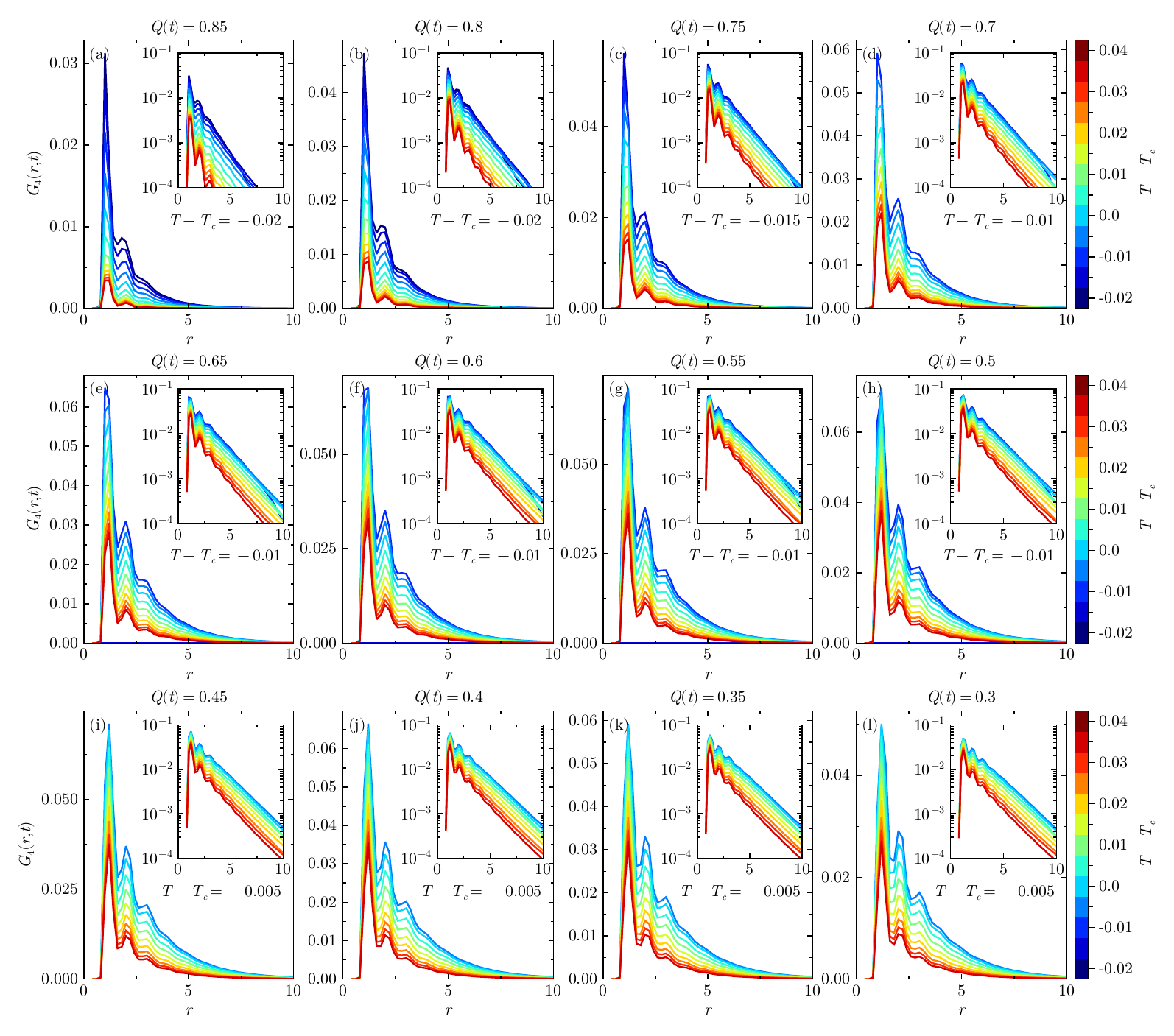}
    \caption{Spatial profiles of the four-point correlation function $G_4(r,t)$ at various intrinsic times and temperatures for a three-dimensional system of $N=$ 100 000 particles. Insets show a semi-logarithmic representation of the same data. The temperature $T-T_c$ indicated in each panel corresponds to the lowest temperature at which the corresponding $Q$-time was resolved.}
    \label{fig:G4_3D_N_100000}
\end{figure*}

\begin{figure*}
    \centering
    \includegraphics[width=\linewidth]{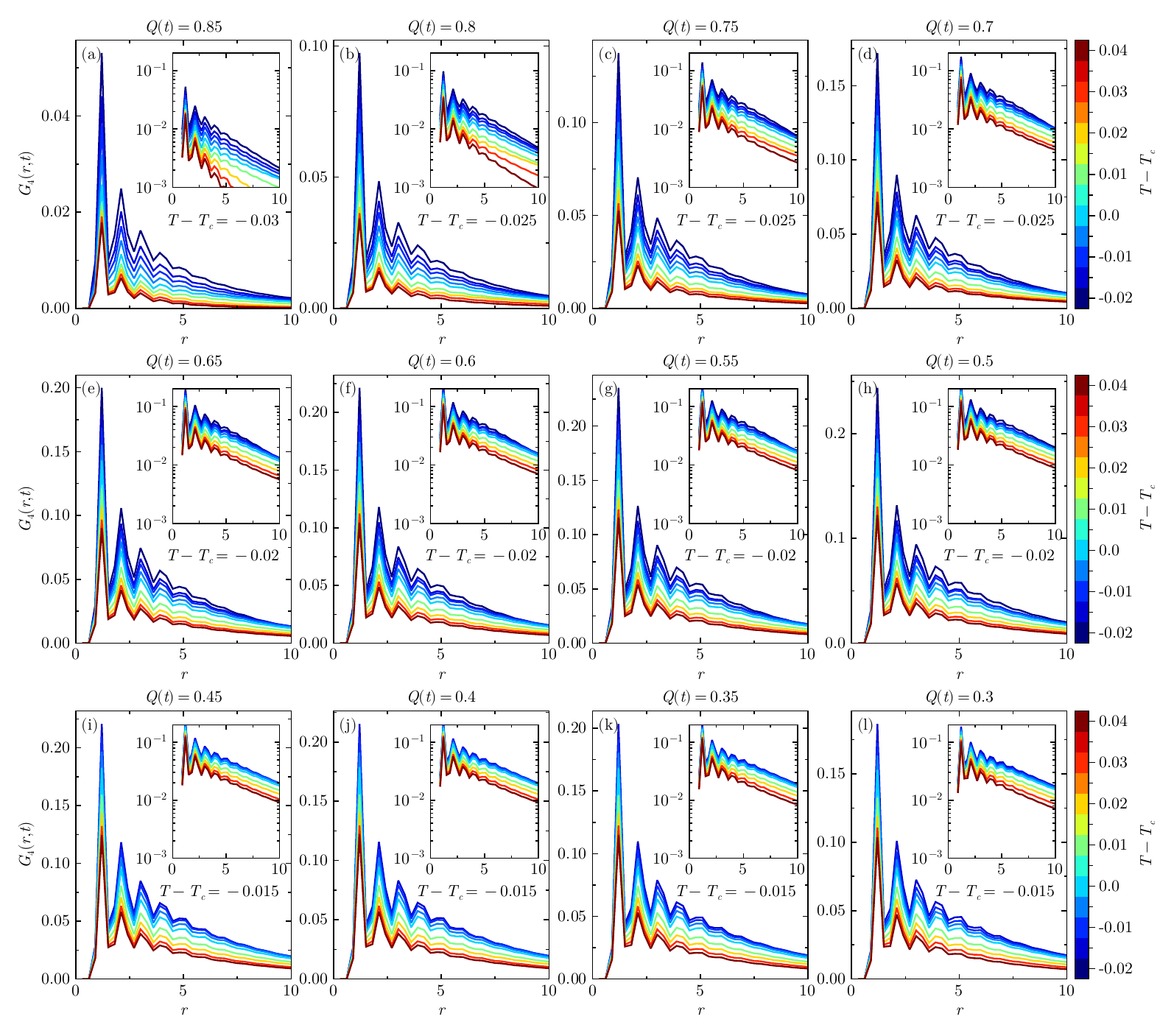}
    \caption{Spatial profiles of the four-point correlation function $G_4(r,t)$ at various intrinsic times and temperatures for a two-dimensional system of $N=$ 10 000 particles. Insets show a semi-logarithmic representation of the same data. The temperature $T-T_c$ indicated in each panel corresponds to the lowest temperature at which the corresponding $Q$-time was resolved.}
    \label{fig:G4_2D_N_10000}
\end{figure*}
As shown in Fig.~1 of the main text, the overlap function $Q(t)$ exhibits a two-step relaxation characteristic of supercooled liquids in both two and three dimensions. At intermediate times, $Q(t)$ plateaus to a value due to transient caging of particles ($\beta$-relaxation). The plateau value $Q_p$, is weakly temperature dependent. At longer times, structural relaxation eventually proceeds, leading to the full decay of $Q(t)$ ($\alpha$-relaxation). Although not shown here, we have checked that our simulations are in quantitative agreement with the relaxation times reported in Ref.~\cite{scalliet2022thirty}, which studied the same system at the same temperatures. 

\subsection{Fluctuations of the Self-Overlap Function} \label{app:Fluctuations of the Self-Overlap Function}
Next, we compute the four-point correlation function introduced in Eq.~2 in the main text (and re-written below)
    \begin{equation}
        G_4(r,t) = \langle [\frac{1}{N}\sum_{i,j=1}^N \delta\mu_i(t)\delta\mu_j(t) \delta(\boldsymbol{r} - \boldsymbol{r}_{ij}(t))]\rangle.
    \label{eq:G4_def_app}
    \end{equation}
where $\delta\mu_i(t) = \mu_i(t) - Q(t)$ denotes the fluctuation of the mobility of
particle $i$ and $\boldsymbol{r}_{ij}(t)$ is the distance vector between particles $i$ and $j$ at time $t$. We note that the definition Eq.~\eqref{eq:G4_def_app} is written differently from earlier studies such as Ref.~\cite{flenner2011analysis}, where instead the separation constraint was evaluated at $t=0$. These are however equivalent by time-reversal and time-translation properties of the equilibrium correlation functions considered here.

At the numerical level, we again compute $G_4(r,t)$ with a time average in a given run and an average over independent realizations. Furthermore, the Dirac $\delta$-function in Eq.~\eqref{eq:G4_def_app} is discretised by histogramming the pair distances into radial bins. Let $\{r_n\}$ denote the centers of bins of
uniform width $\delta r$ and define the indicator function
    \begin{equation}
            \mathcal{I}_n(r) =
        \begin{cases}
          1 & r_n - \frac{\delta r}{2} \leq r < r_n + \frac{\delta r}{2}, \\[0.3em]
          0 & \text{otherwise}.
        \end{cases}    
    \end{equation}
The corresponding shell volume in $d$ spatial dimensions is
    \begin{equation}
        V_n^{(d)} =
        \begin{cases}
          \pi\left((r_n+\frac{\delta r}{2})^2 - (r_n-\frac{\delta r}{2})^2\right), & d=2, \\[0.8em]
          \frac{4\pi}{3}\left((r_n+\frac{\delta r}{2})^3 - (r_n-\frac{\delta r}{2})^3\right), & d=3.
        \end{cases}
    \end{equation}
    The discretised estimator for $G_4(r_{\alpha},t)$ is then
% \begin{widetext}
    \begin{equation}
        G_4(r_{\alpha},t) \approx \frac{1}{V_{\alpha}^{(d)}}\left( \frac{1}{N_{\mathrm{samples}}} \sum_{n=1}^{N_{\mathrm{samples}}}
          \left(\frac{1}{N_{\mathrm{pairs}}} \sum_{i_p=1}^{N_{\mathrm{pairs}}} \frac{1}{N}
         \sum_{i,j=1}^N \delta\mu_i^{(n)}(\Delta t_{i_p})\delta\mu_j^{(n)}(\Delta t_{i_p})
        \mathcal{I}_{\alpha}\big(r_{ij}(\Delta t_{i_p})\big) \right) \right)
    \end{equation} 
% \end{widetext}
which is normalised by the shell volume $V_{\alpha}^{(d)}$. To make it clear, we have $\delta\mu_i^{(n)}(t) = \mu_i^{(n)}(t) - Q(t)$, with $Q(t)$ estimated from Eq.~\eqref{eq:self_overlap_estimator}.

We first demonstrate the validity of the time-average, which, can be seen by the qualitatively (and nearly quantitavely) equivalent profiles of $G_4(r,t)$, shown in Fig.~\ref{fig:comparison_G4_time_averaging_3D_N_10000} for temperatures below the mode-coupling crossover $T_c$. The agreement between the curves also, implicitly, further confirms of our equilibration procedure. Results are shown for a system comprised of $N = 10$ 000 particles in 3D. Analogous behaviour is observed for larger system sizes and dimensions investigated (results not shown). 

\begin{figure*}
    \centering
    \includegraphics[width=\linewidth]{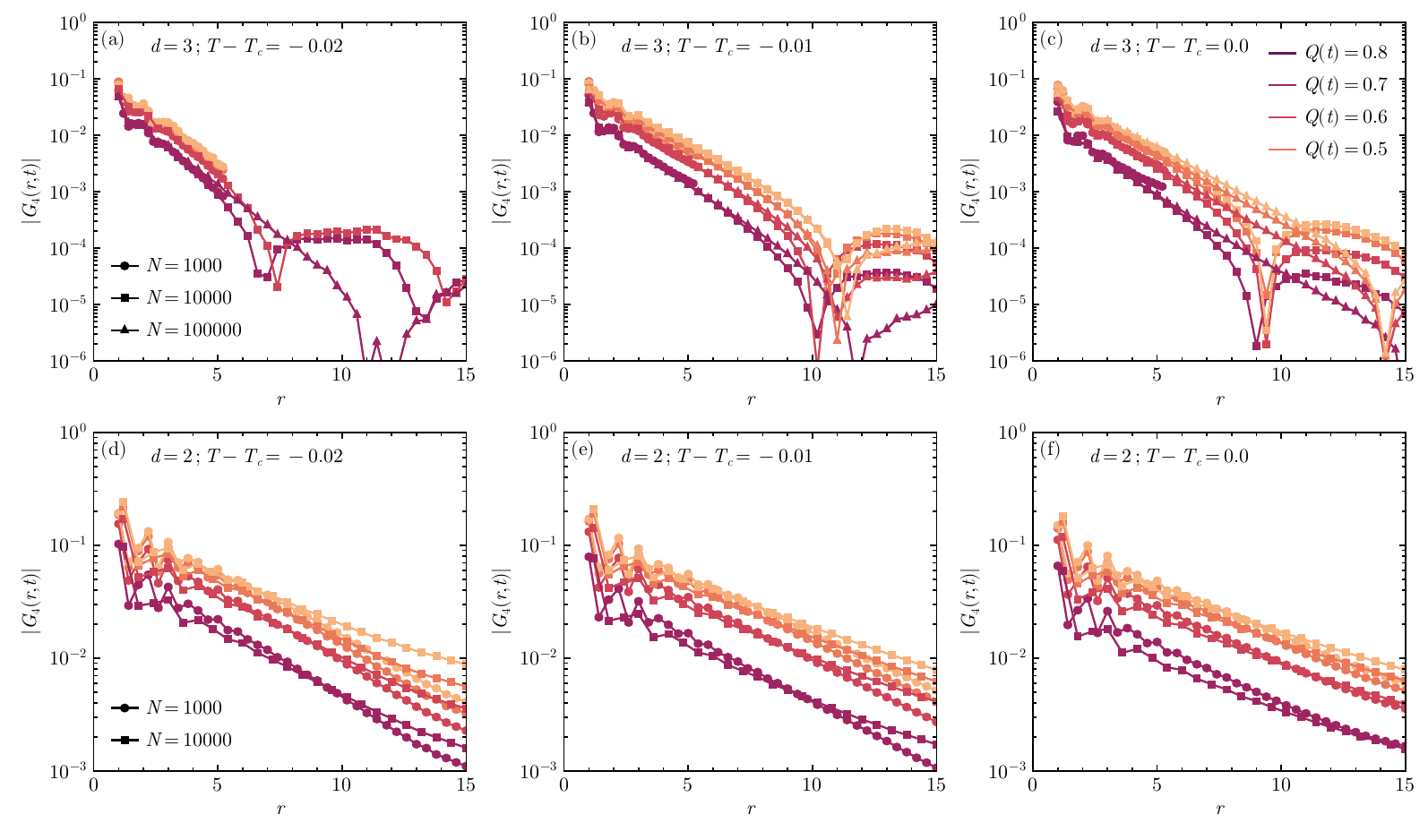}
    \caption{System-size dependence of $|G_4(r,t)|$ at intrinsic time in the $\beta$-regime for the three- (a--c) and two-dimensional (d--f) systems, at temperatures $T - T_c = -0.02$ (a,d), $-0.01$ (b,e), and $0.0$ (c,f). In all cases, the spatial profile at finite $r$ converges with increasing system size, while finite-size artefacts manifest as a negative saturation of $G_4(r,t)$ at large $r$, visible here as a dip due to the absolute value, and shift to larger distances as $N$ increases. The range unaffected by these artefacts, used to define the truncation cutoff $r_c$, is identified as the region where profiles for different system sizes overlap. Note that for the lowest temperature in 3D, we can only resolve down to $Q(t)=0.5$ for the smallest system size, which is why there is only one curve for this $Q$-value.}
\label{fig:compare_sys_size_G4_2x3}
\end{figure*}

We next investigate the system-size dependence of $G_4(r,t)$. The results are shown in Fig.~\ref{fig:compare_sys_size_G4_2x3} (a--c) for 3D and (d--f) for 2D, in which we plot $|G_4(r,t)|$ on a semi-logarithmic scale for temperatures below the crossover temperature ($T \leq T_c$). For the 3D systems, at finite $r$, the correlation profiles exhibit clear convergence as the system size increases, indicating that the local behaviour (up to 5 particle diameters) is already representative of the thermodynamic limit for system sizes of at least $N =$ 10 000 particles. Similar conclusions can be drawn for 2D systems in the $\beta$-regime. The convergence with increasing system size in the $\alpha$-regime at the lowest temperatures is not ideal, but we nevertheless believe that it is sufficient to obtain physically relevant results, as we explain next.

Consistent with this, both the spatial cutoff used in the integration and the range over which we fit the exponential tail lie fully within the region where system-size effects have not yet appeared, further validating our procedure. As expected, the strongest finite-size artefacts shift to progressively larger $r$ as the system size grows, confirming that the large-distance deviations originate from finite simulation volumes rather than from the intrinsic structure of $G_4(r,t)$. Similar convergence behaviour is observed for the higher temperatures as well as the two-dimensional systems investigated shown in Fig.~\ref{fig:compare_sys_size_G4_2x3}. Interestingly, in two dimensions $G_4(r,t)$ sometimes tends to exhibit a saturation of the tail rather than a clean exponential decay. Although the precise origin of this behaviour is not fully understood, it is likely linked to the presence of Mermin–Wagner (MW) fluctuations in the two-dimensional system \cite{illing2017mermin}. Supporting this interpretation, the negative tails expected in the thermodynamic limit reappear when a bond-breaking mobility indicator (immune to MW fluctuations) is used instead (results not shown). 

\begin{figure*}
    \centering
    \includegraphics[width=\linewidth]{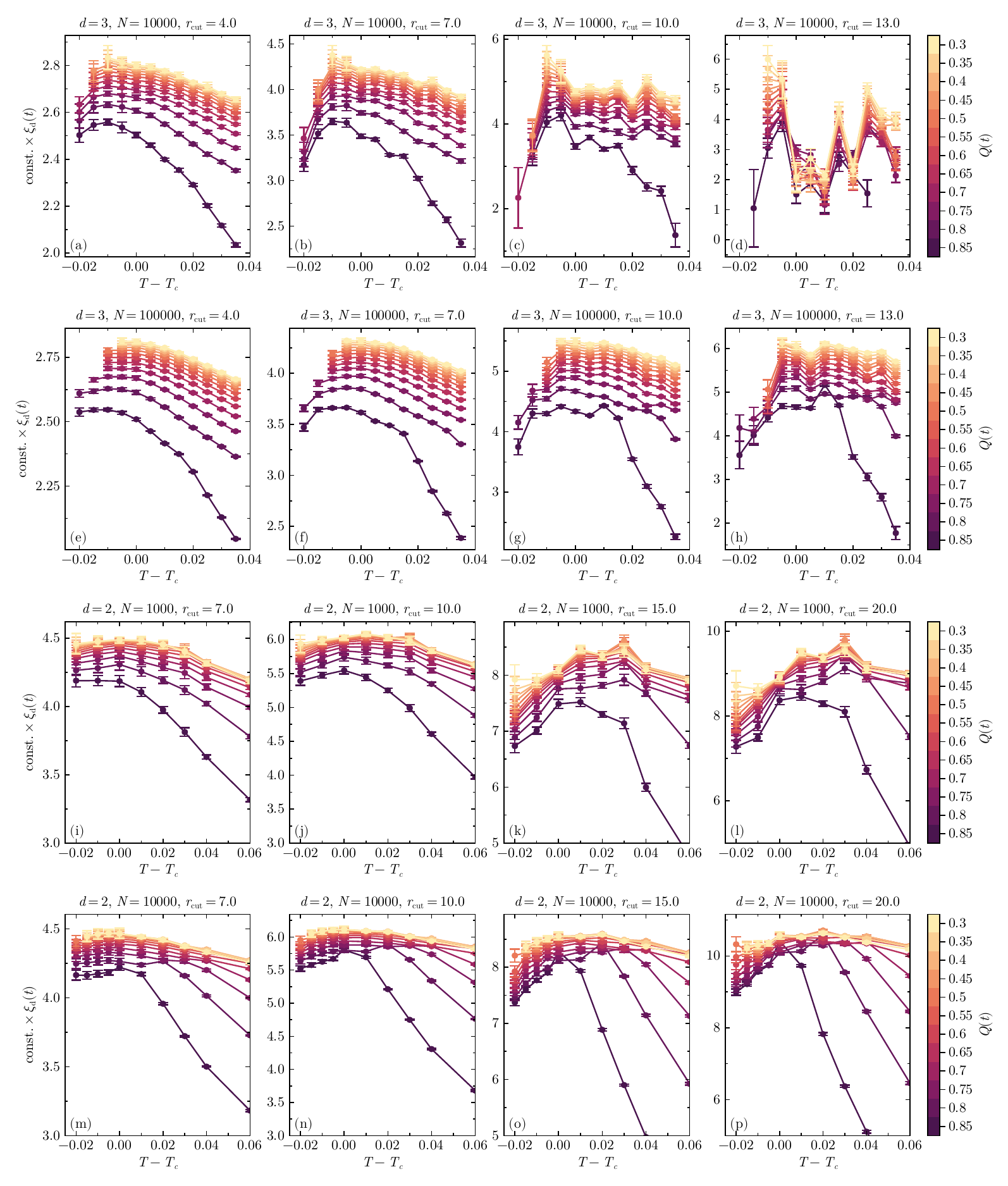}
    \caption{Cut-off dependence of the correlation length $\xi_\mathrm{d}(t)$ obtained from the second-moment method, plotted as a function of $T - T_c$ for several values of the overlap $Q(t)$, for $d = 3$ (top row, $N = 10\,000$, second row, $N = 100\,000$) and $d = 2$ (third row, $N = 1\,000$, bottom row, $N=10\,000$).  Each column corresponds to a different upper integration cut-off $r_c$. As $r_c$ increases, the overall magnitude of $\xi_\mathrm{d}$ grows, yet the non-monotonic temperature dependence remains clearly visible across all cut-offs.  For the largest values of $r_c$ the signal is progressively contaminated by finite-size effects which appear as negative, noisy tails to $G_4(r,t)$, especially for the smaller system size.
    }
\label{fig:xi_vs_Tc_cutoff_dependence}
\end{figure*}

\subsection{The Correlation Length: Additional Results}\label{app:correlation_length}
The correlation length is then computed by considering the Riemann integral of the ratio of the two quantities~:
    \begin{equation}
        \xi_{\mathrm{d}}(t) \approx \mathrm{const} \times \frac{\sum_{\alpha=1}^{N_r} r^2_{\alpha} G_4(r_{\alpha}, t) V_{\alpha}^{(d)} }{\sum_{\alpha=1}^{N_r} G_4(r_{\alpha}, t) V_{\alpha}^{(d)}}
    \end{equation}
with $N_r$ the number of bins in the discretisation of the radial distance and where we compute the second moment as follows 
% \begin{widetext}
    \begin{equation}
        r_{\alpha}^2G_4(r_{\alpha},t) \approx \frac{1}{V_{\alpha}^{(d)}} \left( \frac{1}{N_{\mathrm{samples}}} \sum_{n=1}^{N_{\mathrm{samples}}} \left(\frac{1}{N_{\mathrm{pairs}}} \sum_{i_p=1}^{N_{\mathrm{pairs}}} \left( \frac{1}{N} \sum_{i,j=1}^N \delta\mu_i^{(n)}(\Delta t_{i_p})\delta\mu_j^{(n)}(\Delta t_{i_p})r_{\alpha}^2
    \mathcal{I}_{\alpha}\left(r_{ij}(\Delta t_{i_p})\right) \right)\right) \right).
\end{equation}
% \end{widetext}
Because finite-size effects become significant at large $r$, the integrals are evaluated only up to a cutoff distance $r_c = N_r\delta r$. We have verified that varying $r_c$ within the regime where finite-size effects remain negligible modifies the results only quantitatively and does not affect the trends or conclusions, as we discuss next.

The central observation is that the non-monotonic temperature dependence of $\xi_\mathrm{d}(t)$ is robust across all values of $r_c$ tested, in both dimensions and at all system sizes examined. In particular, the non-monotonicity is already clearly visible at the smallest system sizes ($N = 10\,000$ in $d=3$, $N = 1\,000$ in $d=2$), demonstrating that it is not an artifact of system size or of the precise value of $r_c$.

\begin{figure}
    \centering
    \includegraphics[width=0.75\linewidth]{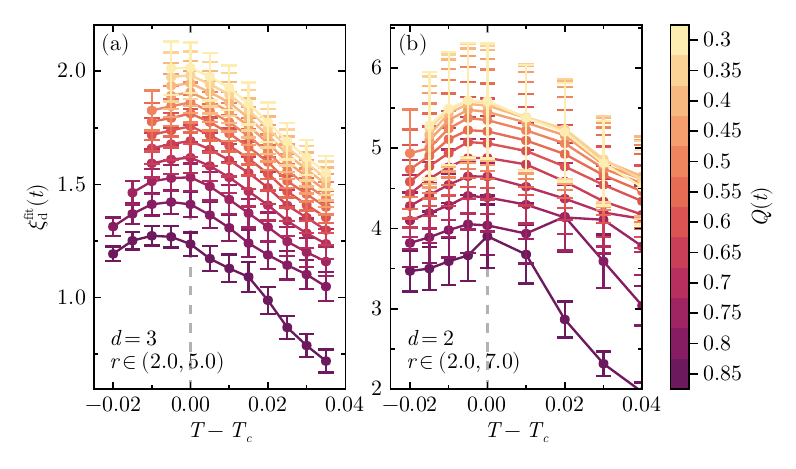}
    \caption{Dynamic correlation length obtained by fitting the exponential tail of $G_4(r,t)$ for the (a) three- and (b) two-dimensional system composed of $N=100$ 000 and $N=10$ 000 particles respectively. The fitting range is indicated in each panels.}
    \label{fig:xi_vs_Tc_fit}
\end{figure}

\begin{figure}
    \centering
    \includegraphics[width=0.6\linewidth]{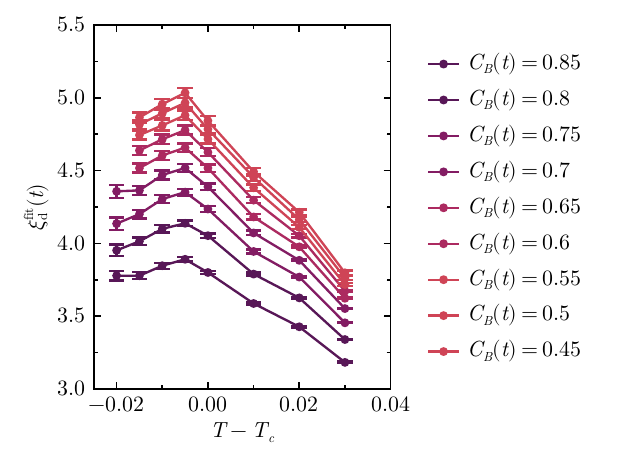}
    \caption{Dynamic correlation length obtained by fitting the exponential tail of $G_4(r,t)$ with a bond-breaking mobility indicator for the two-dimensional system composed of $N=10$ 000 particles. The fitting range is the same as that of Fig.~\ref{fig:xi_vs_Tc_fit}.}
    \label{fig:dynamical_lengthscale_CB}
\end{figure}

As $r_c$ increases, the overall magnitude of $\xi_\mathrm{d}(t)$ grows, since larger cut-offs include contributions from the tails of $G_4(r,t)$ that are excluded at smaller $r_c$. This dependence of the magnitude on $r_c$ is expected and physically unproblematic: the second-moment estimator is sensitive to the range of integration. The temperature dependence and the location of the maximum near $T_c$ are unaffected by this choice.

When $r_c$ is pushed into the regime where finite-size effects become significant, the signal is progressively contaminated by the negative, noisy tails of $G_4(r,t)$. This contamination manifests as increased scatter and erratic behavior in $\xi_\mathrm{d}(t)$. It is most pronounced for the smaller system sizes in $d=3$ (panels c--d), where finite-size effects set in at shorter distances. In $d=3$, increasing the system size from $N=10\,000$ to $N=100\,000$ yields a clear convergence of the magnitude of $\xi_\mathrm{d}(t)$ at intermediate values of $r_c$, confirming that the larger system provides a cleaner signal. Convergence with system size is less complete in $d=2$, consistent with the stronger finite-size effects, more pronounced structural ordering known to affect two-dimensional glass formers~\cite{tong2023emerging} and potentially the effect of Mermin-Wagner fluctuations \cite{illing2017mermin}.

In the main text, we work with values of $r_c$ that exceed $\xi_\mathrm{d}(t)$ by a factor of $1.75$--$2$, verified \emph{a posteriori}, while remaining safely below the regime where finite-size contamination becomes significant. This choice is applied consistently across all temperatures and intrinsic times, and the robustness analysis presented here confirms that it does not affect the qualitative conclusions of the paper. Note that for clarity, the main text only shows curves for $Q(t)=$ 0.8, 0.7, ... 0.3 whereas we show here also the intermediate values $Q(t)=$ 0.8, 0.75, 0.7 ... 0.3. 

We have checked that the non-monotonicity of the correlation length can also be appropriately capture through direct fitting of the exponential tail. The results are shown in Fig.~\ref{fig:xi_vs_Tc_fit}(a)-(b) for the three- and two-dimensional systems respectively. The fitting range is indicated in the plot and was determined to be in the regime where the thermodynamic limit is obeyed. Note that, in the fitting procedure, we fixed the exponent to $p = 0$ in the scaling form given by Eq.~(3) of the main text. In a mean-field limit, it is possible to show that the exponent's dimensional dependence reads $p(d) = - 3/2 + d/2$ and thus $p=0$ is justifiable in three dimensions. Regardless, we expect this term to be sub-leading relative to the exponential tail. We also verified that the correlation length is insensitive to the precise choice of mobility indicator. In Fig.~\ref{fig:dynamical_lengthscale_CB}, we report the fitted dynamical length scale obtained using a bond-breaking mobility indicator \cite{pihlajamaa2025polydispersity} for 2D systems with $N =$ 10 000 particles. The results clearly exhibit the same non-monotonic behaviour as that observed in the main text, demonstrating that this non-monotonicity is robust and unaffected by MW fluctuations.

\section{Numerical Details for SBR}

The SBR equation [Eq.~(9) of the main text] must be solved numerically on a finite spatial grid. In this setting, spatial self-averaging is incomplete, and disorder-averaged observables are instead obtained by averaging over many independent realizations of $\delta s(\boldsymbol{x})$ drawn from the prescribed distribution. All SBR results presented below are obtained in this way, with the number of realizations chosen to ensure convergence of the disorder average.

The SBR parameters for the 3D and 2D cases are reported in Table~\ref{tab:SBR_params}. In both cases, the exponent parameter $\lambda$ is computed from the dynamical exponent $\gamma$ provided in Ref.~\cite{scalliet2022thirty} for our model system, while the remaining parameters are fitted to best reproduce $\xi_{\mathrm{d}}(t)$. The exponent parameter $\lambda$ is defined by
    \begin{equation}
        \lambda = \frac{\Gamma(1-a)^2}{\Gamma(1-2a)} = \frac{\Gamma(1+b)^2}{\Gamma(1+2b)},
    \end{equation}
where $a$ and $b$ are the critical exponents governing the power-law relaxation on the $\beta$-relaxation timescale, related to the divergence exponent $\gamma$ of the relaxation time via $\gamma = 1/(2a) + 1/(2b)$ \cite{gotze2009complex}. Given $\gamma$, this system of equations determines $\lambda$ and the exponents $a$ and $b$ self-consistently. The SBR equations are solved on a cubic grid of $N=40^3$ points ($d=3$) and a square grid of $N=200^2$ points ($d=2$), both with periodic boundary conditions, using a logarithmically coarse-grained time 
integration scheme.

As mentioned in the main text, the remaining parameters are fitted to best reproduce the dynamic correlation length measured in the simulations.
In order to do this we have exploited the fact that the only parameters in the SBR equations are $\lambda$ and the (adimensional) reduced temperature $\sigma$, $\Delta \sigma^{\frac{8}{d-8}}$ and $\alpha^{\frac{2d}{8-d}}$. Indeed, according to the discussion in Sec. VI of the supplemental material of \cite{rizzo2020solvable}
 the SBR solutions for given values of $\alpha$, $\Delta \sigma$ and $\sigma$ are related by simple rescalings $b_\phi$, $b_x$ and $b_{\sigma}$ to a {\it universal} solution depending only on the two adimensional parameters. Thus we first obtained numerically the universal solution and then tuned  the parameters $b_{\phi}$, $b_x$ and $b_{\sigma}$  in order to fit the numerical data. Then the parameters $\alpha$, $\Delta$ and $\sigma$ were obtained inverting equations (50-52) in Sec. VI of the supplemental material of \cite{rizzo2020solvable}.

Next we discuss how we obtain the correlation lengths from the numerical solution.
We have checked that $G^{\mathrm{SBR}}_4(r,t)$ decays with a pure exponential $G^{\mathrm{SBR}}_4(r,t) \propto e^{r/\xi_d}$ in 3D and with $G^{\mathrm{SBR}}_4(r,t) \propto r^{1/2} \, e^{r/\xi_d}$ in 2D consistently with the  violation of Ornstein-Zernike theory in mean-field due to the presence of the quenched disorder. Indeed, as discussed in Sec. V of the Supplemental Materials of \cite{rizzo2020solvable}, in mean-field the Fourier transform of $G^{\mathrm{SBR}}_4(r,t)$ is a {\it squared} Ornstein-Zernike form $(k^2+\xi^{-2})^{-2}$ that leads to the aforementioned behavior $p(d) = - 3/2 + d/2$.
In practice the function $G^{\mathrm{SBR}}_4(r,t)$ is obtained numerically with $r= n \Delta x$ and $n=0,1,2,\dots$.  $G^{\mathrm{SBR}}_4(r,t)$ follows the exponential decay at large enough $r$ and then plateaus to a value that decreases with the grid size. Therefore the exponential behavior is visible for {\it large but not too large} values of $r$.
For all the times and temperatures reported we have checked that in 3D this occurs for values of $r$ such that $G^{\mathrm{SBR}}_4(r,t)$ has decayed to one tenth of its $r=0$ value $G^{\mathrm{SBR}}_4(0,t)$.

More precisely we have considered the rescaled function $\tilde{G}(r)=G^{\mathrm{SBR}}_4(r)/G^{\mathrm{SBR}}_4(0)$ where we have omitted the explicit time dependence. Then the point $r^*$ was determined such that  $\tilde{G}(r^*+dx)>0.1>\tilde{G}(r^*)$ and the function $\ln \tilde{G}(r)$ was fitted with a linear fit on the two points $r^*+dx$ and $r^*$.
Due to the finite grid spacing, the point $r^*$ takes values in $n\, \Delta x$ and therefore it makes intermittent small  jumps $\Delta x$ upon changing the time $t$ and/or the temperature. In correspondence of these jumps the correlation length estimated from the fit between $r^*+dx$ and $r^*$ also displays a small jump. In order avoid this effect we considered a linear fit $f_{\mathrm{min}}$ of  $\ln \tilde{G}(r)$ between $r^*+dx$ and $r^*$ and a linear fit $f_{\mathrm{max}}$ of  $\ln \tilde{G}(r)$ between $r^*+2\,dx$ and $r^*+\Delta x$. The two fits are then summed with weights inversely proportional respectively to the distance between $0.1$ and $\tilde{G}(r^*)$ ($d_{\mathrm{min}} \equiv 0.1-\tilde{G}(r^*)$) and to the distance between $\tilde{G}(r^*+dx)$ and  $0.1$, ($d_{\mathrm{max}}=\tilde{G}(r^*+dx)-0.1$): $f= (f_{\mathrm{min}}\, d_{\mathrm{max}}+f_{\mathrm{max}}d_{\mathrm{min}})/(d_{\mathrm{max}}+d_{\mathrm{min}})$. This procedure grants that the fitted parameters are continuous also when $r^*$ jumps abruptly by a quantity $\pm \Delta x$ upon an infinitesimal change of the time and/or the temperature.
Once the correlation length has been obtained from the exponentail decay we can also compute the correlation length with the alternate definion based on the second moment. To do this it is essential to perform the numerical integral over $\tilde{G}(r)$ extrapolating the function beyond $r^*$ by the exponential fit, the region of large $r$ is indeed greatly enhanced by the factor $r^4$ at the numerator and by the factor $r^2$ and the denominator. 
A similar procedure has been used for the 2D case.

\begin{table}
\centering
\begin{tabular}{||l|c|c||}
\hline
Spatial Dimension & $d=3$ & $d=2$ \\
\hline
Critical Temperature $T_c$ & $0.117$ & $0.130$ \\
\hline
Plateau Value $Q_p$ & $0.856$ & $0.800$ \\
\hline
Exponent Parameter $\lambda$ & $0.74$ & $0.78$ \\
\hline
Reduced Temp. prefactor $\sigma/(T_c - T)$ & $-0.0328$ & $-0.125$ \\
\hline
Stiffness $\alpha$ & $0.373$ & $0.0562$ \\
\hline
Disorder Strength $\Delta \sigma$ & $0.0391$ & $0.00832$ \\
\hline
Grid Spacing $\Delta x$ & $2.4$ & $0.375$ \\
\hline
\end{tabular}
\caption{SBR parameters used to generate the theoretical results in 
Fig.~5 of the main text. The exponent parameter $\lambda$ is taken from \citet{scalliet2022thirty} for the same model system; all other parameters are fitted to best reproduce $\xi_{\mathrm{d}}(t)$.}
\label{tab:SBR_params}
\end{table}

\end{document}